\documentclass[final,3p,times]{elsarticle}
\usepackage{amssymb}
\biboptions{sort&compress}
\usepackage{amsfonts}
\usepackage{amssymb}
\usepackage{amsmath}
\usepackage{graphics}
\usepackage{graphicx}
\usepackage{subfigure}
\usepackage{epsfig}
\usepackage{epstopdf}
\usepackage{amsthm}
\usepackage{textcomp}
\usepackage{color}
\usepackage{float}
\usepackage{multirow}
\usepackage{float}
\usepackage{graphicx}
\usepackage{subfigure}
\usepackage{amsmath,amssymb,amsthm}
\usepackage{amsmath}
\usepackage{amssymb}
\usepackage{epstopdf}
\usepackage{longtable}
\usepackage{rotating}
\usepackage{multirow}
\usepackage{mathrsfs}

\journal{Communications in Nonlinear Science and Numerical Simulation}
\begin{document}
\begin{frontmatter}



\cortext[cor1]{Corresponding author}
%%\cortext[cor2]{Principal corresponding author}
\title{A discount strategy in word-of-mouth marketing and its assessment}


\author[label1]{Tianrui Zhang}
\ead{363726657@qq.com}

\author[label1]{Xiaofan Yang}
\ead{xfyang1964@gmail.com}

\author[label1,label2]{Lu-Xing Yang\corref{cor1}}
\ead{ylx910920@gmail.com}

\author[label3]{Yuan Yan Tang}
\ead{yytang@umac.mo}

\author[label1]{Yingbo Wu}
\ead{wyb@cqu.edu.cn}

\address[label1]{College of Software Engineering, Chongqing University, Chongqing, 400044, China}

\address[label2]{Faculty of Electrical Engineering, Mathematics and Computer Science, Delft University of Technology, Delft, GA 2600, The Netherlands}

\address[label3]{Department of Computer and Infomation Science, The University of Macau, Macau}

\begin{abstract}
This paper addresses the discount pricing in word-of-mouth (WOM) marketing. A new discount strategy known as the Infection-Based Discount (IBD) strategy is proposed. The basic idea of the IBD strategy lies in that each customer enjoys a discount that is linearly proportional to his/her influence in the WOM network. To evaluate the performance of the IBD strategy, the WOM spreading process is modeled as a dynamic model known as the DPA model, and the performance of the IBD strategy is modeled as a function of the basic discount. Next, the influence of different factors, including the basic discount and the WOM network, on the dynamics of the DPA model is revealed experimentally. Finally, the influence of different factors on the performance of the IBD strategy is uncovered experimentally. On this basis, some promotional measures are recommended.
\end{abstract}


\end{frontmatter}

% \linenumbers


\section{Introduction}

In all ages, discount pricing, whereby  a price is listed as a discount from an earlier or regular price, has been accepted as a common marketing practice. As opposed to a merely low price, a discounted price can make a rational consumer more willing to purchase the item, because (1) the information that the product was initially sold at a high price can indicate that the product is high quality, and (2) a discounted price can signal that the product is an unusual bargain, and there is little point for lower prices \cite{Monahan1984, Lal1984, Lee1986, Shah2005, Armstrong2013}.

Word-of-mouth (WOM), which is loosely defined as the sharing of information about a product between a consumer and his/her friends, colleagues, and acquaintances, has long been a prevalent topic among marketing practitioners and researchers \cite{Dichter1966}. As compared to the traditional advertising, WOM plays a more crucial role in influencing consumer decisions \cite{Katz1966}. For instance, customers linked to a prior customer will be 3 to 5 times more likely to purchase the product \cite{Hill2006}. Due to significantly low cost and rapid spreading, WOM outperforms advertising in terms of marketing profit \cite{Misner1999,Chevalier2006}. With the popularity of online social networks such as Facebook and Twitter, WOM has been turned into the main measure of marketing \cite{Trusov2009}.

The major concern on WOM marketing is the influence maximization problem: find a set of seeds such that the expected number of individuals activated from this seed set is maximized \cite{Peres2010}, and large numbers of seeding algorithms have been proposed \cite{Kempe2005, ChenW2009, Mochalova2014, Kempe2015, ZhangHY2016b}. To evaluate the performance of different seeding algorithms, a number of dynamic models capturing the WOM spreading process have been suggested \cite{Bass1969, YuY2003, WeiXT2013, Gardner2013, Sohn2013, LiS2013, LiS2014, Rodrigues2015, JiangP2017}. As it is known, the structure of the WOM network has a significant influence on the marketing performance \cite{Bampo2008, Siri2013}. The above dynamic models are all population-level, which do not take into account the structure of the WOM network. As thus, these models are incompetent to analyze the influence of the WOM network on the marketing profit. To uncover the impact, the WOM spreading process must be captured more exactly by borrowing the network-level modeling technique \cite{Satorras2011, Castellano2010, YangLX2014, Satorras2015, YangLX2017a} or the individual-level modeling technique \cite{Mieghem2009, Sahneh2013, YangLX2015, XuSH2015, YangLX2017b, YangLX2017c} in the epidemic dynamics.

It sounds a good idea to take a proper discount strategy in WOM marketing, because the discount aided with the WOM may greatly enhance the marketing profit. In this context, it is necessary to evaluate the performance of different discount strategies with the aid of individual-level WOM spreading models. As far as we know, hitherto there has been no literature in this aspect.

This paper addresses the discount pricing in WOM marketing. A discount strategy known as the Infection-Based Discount (IBD) strategy, in which each customer enjoys a discount that is linearly proportional to his/her influence in the WOM network, is proposed. To evaluate the performance of the IBD strategy, the WOM spreading process is modeled as a dynamic model known as the DPA model, and the performance of the IBD strategy is modeled as a function of the basic discount. Next, the influence of different factors, including the basic discount and the WOM network, on the dynamics of the DPA model is revealed experimentally. Finally, the influence of different factors on the performance of the IBD strategy is uncovered experimentally. On this basis, some promotional measures are recommended.

The subsequent materials are organized in this fashion. Section 2 describes the new discount strategy. Section 3 models the WOM spreading process and the performance of the discount strategy, respectively. Section 4 experimentally reveals the impact of different factors on the dynamics of the WOM spreading model. Section 5 experimentally uncovers the influence of different factors on the performance of the discount strategy. Finally, this work is closed by Section 6.

\section{A new discount strategy}

Given a large collection of products to be promoted in a WOM marketing campaign, and given the target market for the campaign, this section is intended to design a discount strategy.

As compared to a less influential prior customer, a more influential prior customer in the WOM network has a stronger influence on customers' purchase decisions and, hence, bring about a higher profit for the merchant. Inspired by this idea, we describe a discount strategy as follows. \\

\emph{The discount given to each customer is linearly proportional to his/her influence in the WOM network.}\\

We refer the discount strategy as the \emph{influence-based discount} strategy or simply the IBD strategy. Next, let us formalize the discount strategy. Suppose the target market consists of $N$ individuals labeled as $1, 2, \cdots, N$. Let $G = (V, E)$ denote the WOM network, where $V = \{1, 2, \cdots, N\}$, and $(i,j) \in E$ if and only if customer $j$ can recommend products to customer $i$. Let $\mathbf{A} = \left(a_{ij}\right)_{N \times N}$ denote the adjacency matrix of $G$. Define the \emph{influential degree} of customer $i$ in the WOM network as
\[
d_i = \frac{\sum_{j=1}^Na_{ji}}{\max_{1 \leq k \leq N}\sum_{j=1}^Na_{jk}}.
\]
Then $0 \leq d_i \leq 1$, and $d_i$ measures the influence of customer $i$ in the WOM network. On this basis, the IBD strategy can be formalized as follows. \\

\emph{For each customer $i$, the discount given to him/her is $\theta d_i$, where $0 \leq \theta \leq 1$, $\theta$ is referred to as the basic discount}.\\

Clearly, the performance of the IBD strategy, which will be evaluated in the subsequent sections, is dependent upon the basic discount as well as the structure of the WOM network.

\section{A mathematical framework for evaluating the performance of the IBD strategy}

This section aims to provide a mathematical framework for evaluating the performance of the IBD strategy proposed in the previous section.

\subsection{The basic notions, notations and hypotheses}

For our purpose, let us introduce some notions, notations and hypotheses as follows.

Suppose the WOM marketing campaign starts at time $t = 0$ and terminates at time $t = T$. At any time, every customer in the target market is assumed to be in one of three possible states: \emph{dormant}, \emph{potential}, and \emph{adopting}. Dormant customers are those who have no will of buying an item, potential customers are those who have the will of buying an item, and adopting customers are those who buy an item. For $1 \leq i \leq N$, let $X_i(t)$ = 0, 1, and 2 denote that at time $t$ customer $i$ is dormant, potential, and adopting, respectively. For $1 \leq i \leq N$, let
\[
D_i(t) = \Pr\{X_i(t) = 0\}, \quad P_i(t) = \Pr\{X_i(t) = 1\}, \quad A_i(t) = \Pr\{X_i(t) = 2\}.
\]
As $D_i(t)+P_i(t)+A_i(t) \equiv 1$, the vector
\[
\mathbf{x}(t) = \left(P_1(t), \cdots, P_N(t), A_1(t), \cdots, A_N(t)\right)^T
\]
probabilistically captures the state of the target market at time $t$. Next, let us impose a collection of hypotheses as follows.

\begin{enumerate}
		
	\item[(H$_1$)] Each customer is only allowed to buy one item per time.
	
	\item[(H$_2$)] (Original price) The original price per item is one unit.
	
	\item[(H$_3$)] (Discount) The discount per item given to a potential customer $i$ is $\theta d_i$, where the basic discount $\theta$ is under control.
	
	\item[(H$_4$)] (WOM) Due to recommendations by adopting customers, at time $t$ a dormant customer $i$ has the will of buying an item and hence becomes potential at rate $\alpha\sum_{j=1}^N a_{ij}A_j(t)$, where $\alpha > 0$ is referred to as the \emph{WOM force}.
	
	\item[(H$_5$)] (Rigid demand) Due to the rigid demand, at time $t$ a potential customer $i$ buys an item and hence becomes adopting at rate $\beta_1$, where $\beta_1 > 0$ is referred to as the \emph{rigid demand}.
	
	\item[(H$_6$)] (Lure) Due to the lure of discount, at time $t$ a potential customer $i$ buys an item and hence becomes adopting at rate $\beta_2 \theta d_i$, where $\beta_2 > 0$ is referred to as the \emph{lure force}.
	
	\item[(H$_7$)] (Viscosity) At any time an adopting customer temporarily loses the will of buying another item and hence becomes dormant at rate $\gamma > 0$, where $\gamma$ is referred to as the \emph{viscosity}.
	
\end{enumerate}

Hypothesis (H$_3$) declares that a more influential node is given a higher discount than a less influential node. (H$_4$) implies that a more influential node contributes more to the marketing than a less influential node. Fig. 1 shows hypotheses (H$_4$)-(H$_7$) schematically.

\begin{figure}[H]
	\centerline{\includegraphics[width=8cm]{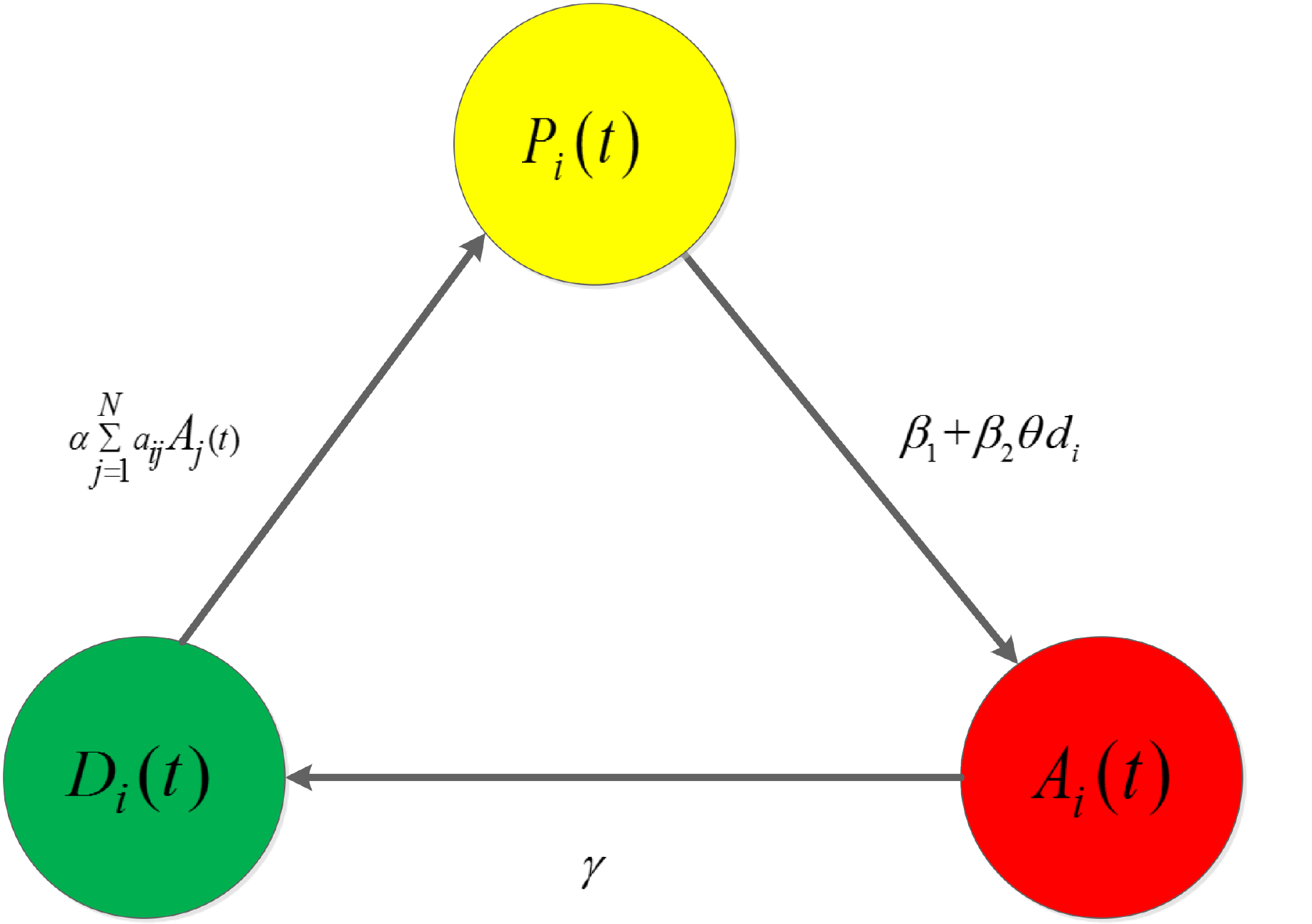}}
	\caption{Diagram of hypotheses (H$_4$)-(H$_7$).}
\end{figure}

\subsection{The DPA model}

WOM spreading plays a key role in the IBD strategy. Starting from the basic hypotheses, let us derive the dynamic model capturing the WOM spreading process.

Let $\Delta t$ be a very small time interval. Hypotheses (H$_4$)-(H$_7$)
imply that, for $1 \leq i \leq N$,
\[
\begin{split}
\Pr\{X_i(t + \Delta t) = 1  \mid X_i(t) = 0\}
&= \alpha \Delta t \sum_{j=1}^N a_{ij}A_j(t) +o(\Delta t), \\
\Pr\{X_i(t + \Delta t) = 2  \mid X_i(t) = 0\} &= o(\Delta t), \\
\Pr\{X_i(t + \Delta t) = 0  \mid X_i(t) = 1\} &= o(\Delta t), \\
\Pr\{X_i(t + \Delta t) = 2  \mid X_i(t) = 1\} &= (\beta_1 + \beta_2 \theta d_i) \Delta t+o(\Delta t), \\
\Pr\{X_i(t + \Delta t) = 0  \mid X_i(t) = 2\} &= \gamma \Delta t + o(\Delta t), \\
\Pr\{X_i(t + \Delta t) = 1  \mid X_i(t) = 2\} &= o(\Delta t). \\
\end{split}
\]
As a result, we have
\[
\begin{split}
\Pr\{X_i(t + \Delta t) = 0  \mid X_i(t) = 0\}
&= 1 - \alpha \Delta t \sum_{j=1}^N a_{ij}A_j(t) +o(\Delta t), \\
\Pr\{X_i(t + \Delta t) = 1  \mid X_i(t) = 1\} &= 1 - (\beta_1 + \beta_2 \theta d_i) \Delta t+o(\Delta t), \\
\Pr\{X_i(t + \Delta t) = 2  \mid X_i(t) = 2\} &= 1 - \gamma \Delta t + o(\Delta t). \\
\end{split}
\]
By the total probability formula, we have, for $1 \leq i \leq N$,
\[
\begin{split}
P_i(t + \Delta t) &=  D_i(t) \Pr\{X_i(t + \Delta t) = 1 \mid X_i(t) = 0\} + P_i(t) \Pr\{X_i(t + \Delta t) = 1 \mid X_i(t) = 1\} \\
& \quad + A_i(t) \Pr\{X_i(t + \Delta t) = 1 \mid X_i(t) = 2\} \\
&= \alpha \Delta t[1 - P_i(t) - A_i(t)]\sum_{j=1}^Na_{ij}A_j(t) + P_i(t)\left[1 - (\beta_1 + \beta_2\theta d_i)\Delta t\right] + o(\Delta t),
\end{split}
\]
\[
\begin{split}
A_i(t + \Delta t) &= D_i(t) \Pr\{X_i(t + \Delta t) = 2 \mid X_i(t) = 0\} + P_i(t) \Pr\{X_i(t + \Delta t) = 2 \mid X_i(t) = 1\} \\
& \quad + A_i(t) \Pr\{X_i(t + \Delta t) = 2 \mid X_i(t) = 2\} \\
& = P_i(t)(\beta_1 + \beta_2 \theta d_i)\Delta t + A_i(t)(1 - \gamma \Delta t) + o(\Delta t).
\end{split}
\]
Rearranging the terms, dividing both sides by $\Delta t$, and letting $\Delta t \rightarrow 0$, we get the following dynamical model.
\begin{equation*}
	\left\{
	\begin{aligned}
		\frac{dP_i(t)}{dt}&= \alpha \left[1-P_i(t)-A_i(t)\right]\sum_{j=1}^N a_{ij}A_j(t)  - (\beta_1 + \beta_2 \theta d_i) P_i(t),
		\quad 1 \leq i \leq N,\\
		\frac{dA_i(t)}{dt}&= (\beta_1 + \beta_2 \theta d_i) P_i(t) - \gamma A_i(t),
		\quad 1 \leq i \leq N.
	\end{aligned}
	\right.
\end{equation*}
We refer to this model as the \emph{DPA model}. The model can be rewritten in matrix notion as $\frac{d\mathbf{x}(t)}{dt} = \mathbf{f}(\mathbf{x}(t))$.

Clearly, the dynamics of the DPA model is dependent on the WOM force, the rigid demand, the lure force, the viscosity, the WOM network, and the basic discount.

\subsection{The expected profit}

The performance of the IBD strategy can be measured by its expected profit; the higher the expected profit, the better the IBD strategy. In view of hypotheses (H$_1$)-(H$_3$) and according to the DPA model, the expected profit of the IBD strategy can be measured by
\[
EP(\theta) = \int_0^T \sum_{i=1}^N (\beta_1 + \beta_2 \theta d_i) P_i(t)(1 - \theta d_i)dt.
\]

Clearly, the expected profit is dependent on the WOM force, the rigid demand, the lure force, the viscosity, the WOM network, and the basic discount.

\subsection{The WOM network}

For the purpose of investigating the influence of different factors on the dynamics of the DPA model and the profit gained with the IBD strategy, let us build six candidate WOM networks. Fig. 2 depicts three small-world networks \cite{Watts1998}, and Fig. 3 exhibits three scale-free networks \cite{Barabasi1999}.

\begin{figure}[H]
	\subfigure[$SW_1$]{\includegraphics[width=0.33\textwidth,height=4cm]{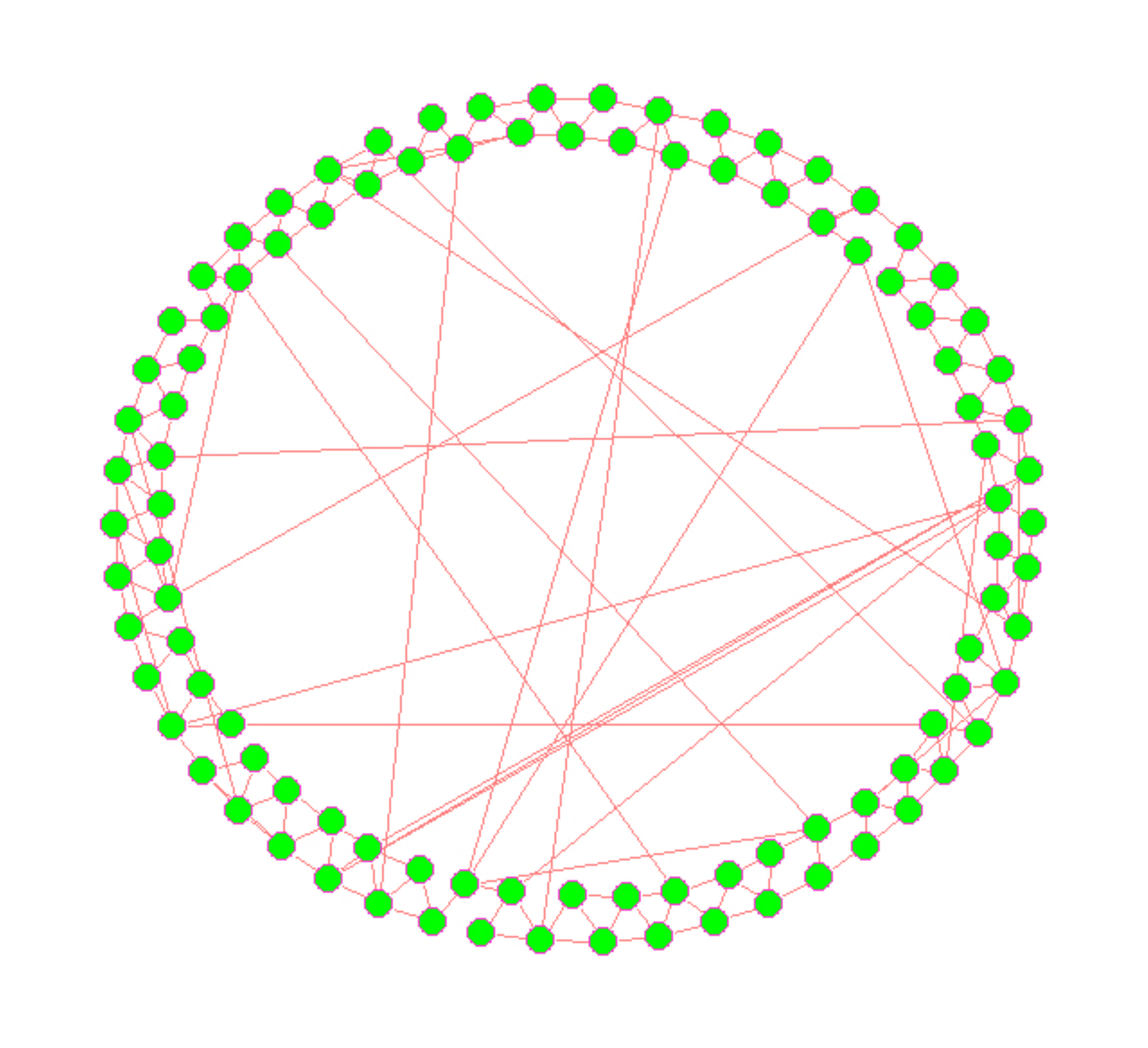}\label{a}}
	\subfigure[$SW_2$]{\includegraphics[width=0.33\textwidth,height=4cm]{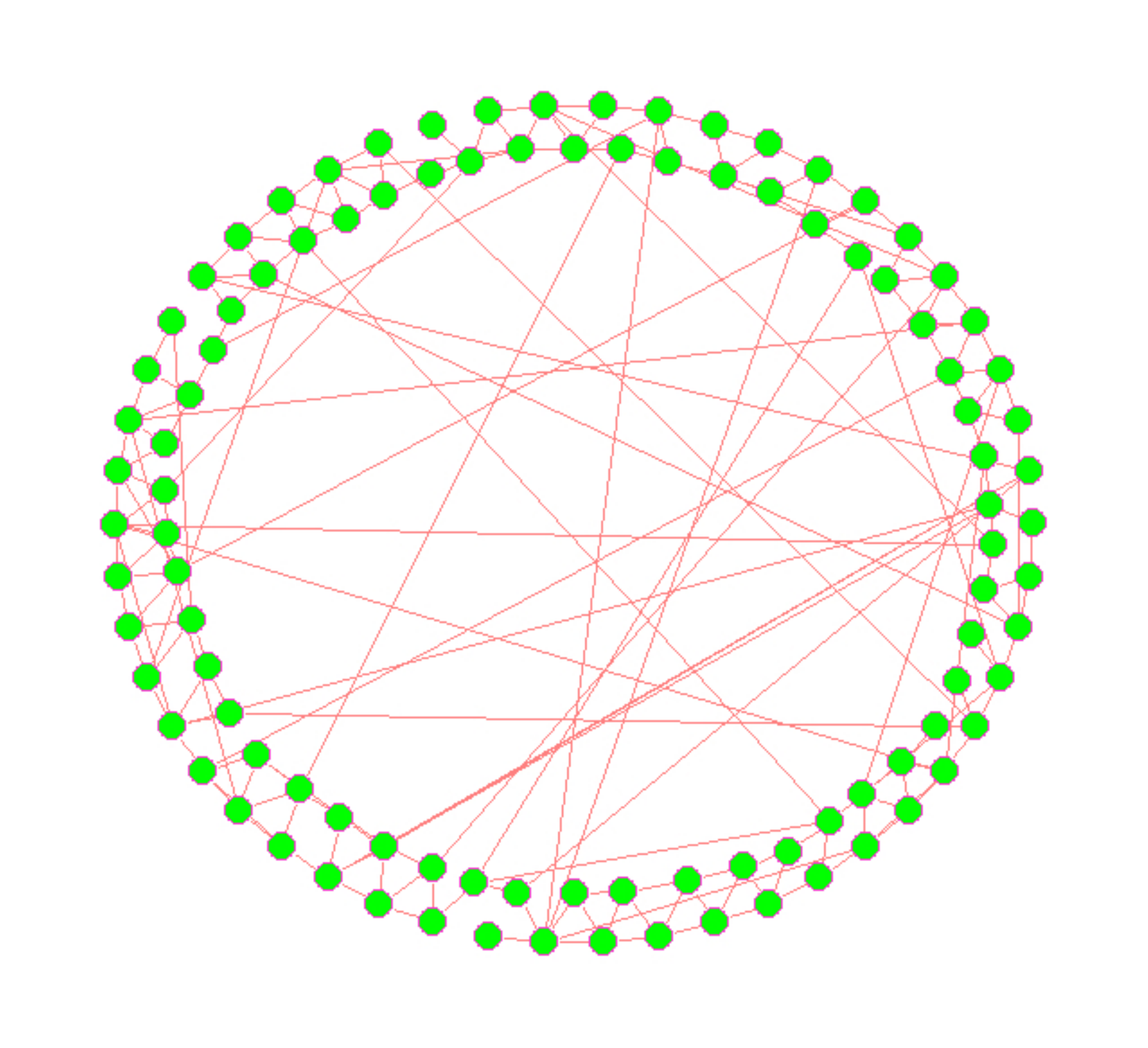}\label{a}}
	\subfigure[$SW_3$]{\includegraphics[width=0.33\textwidth,height=4cm]{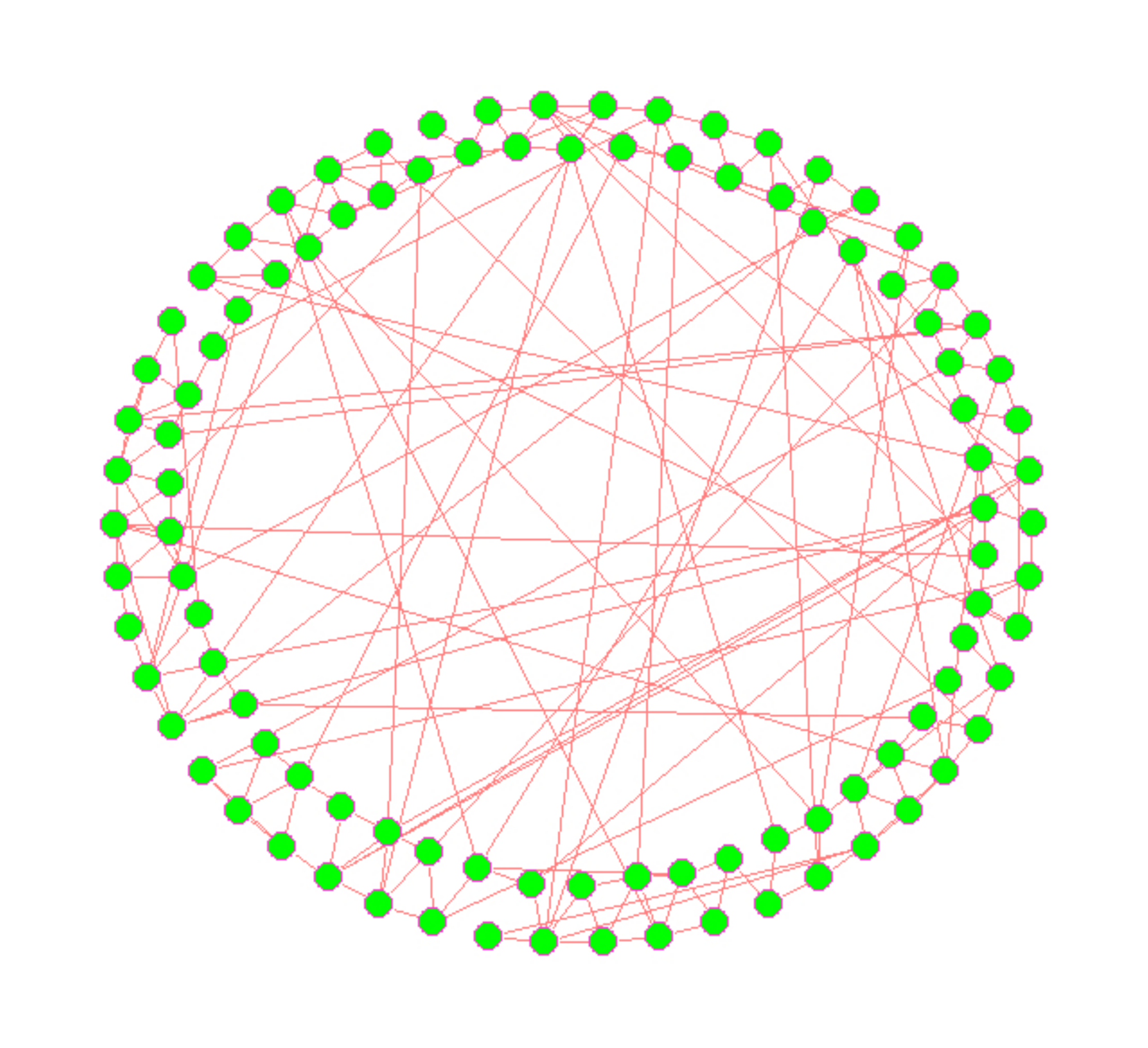}\label{a}}
	\vspace{-3ex}
	\caption{Three small-world networks with 100 nodes, 200 edges, and edge-rewiring probability $p = 0.1, 0.2, 0.3$, respectively.}
\end{figure}

\begin{figure}[H]
	\subfigure[$SF_1$]{\includegraphics[width=0.33\textwidth,height=4cm]{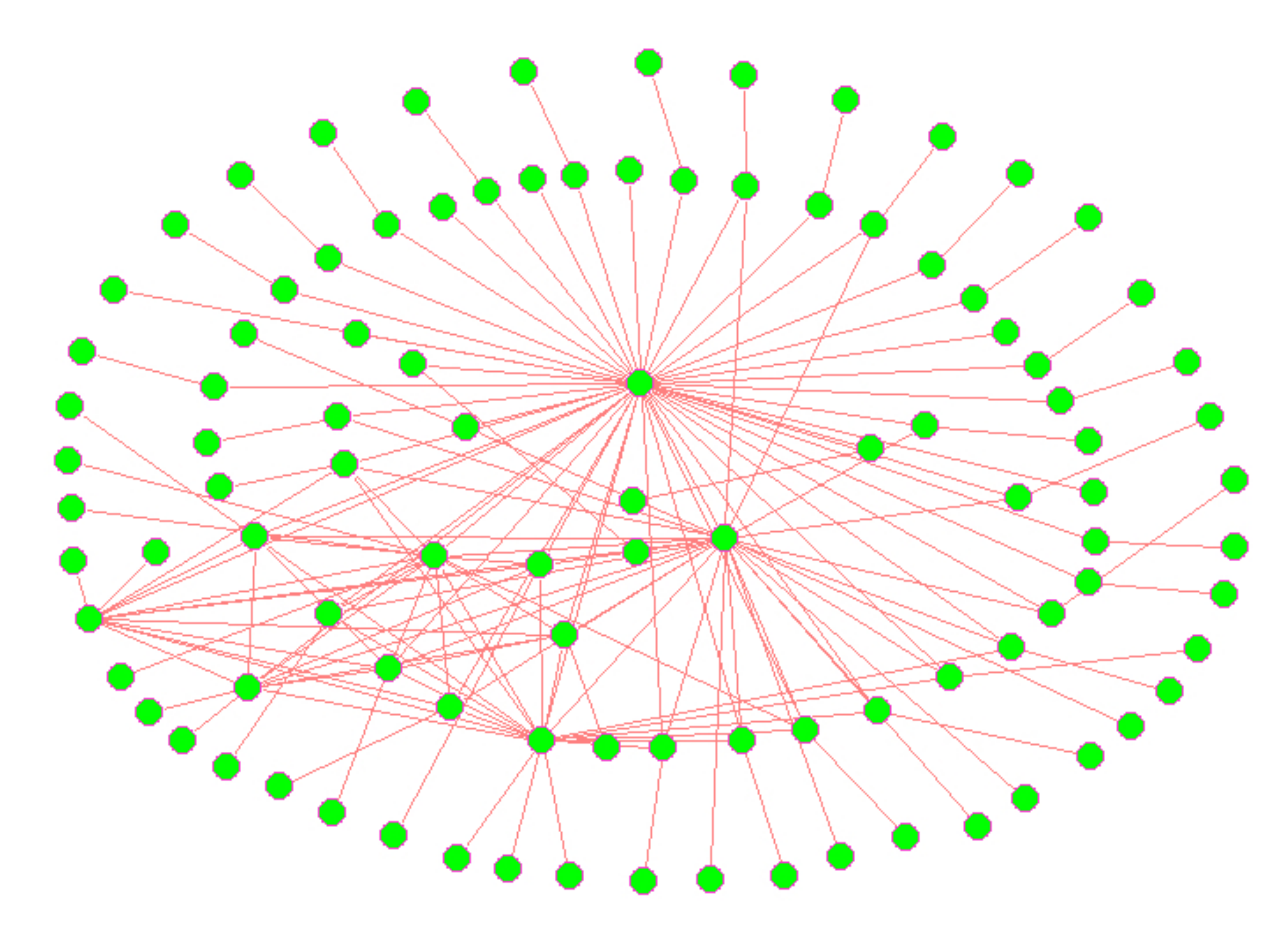}\label{a}}
	\subfigure[$SF_2$]{\includegraphics[width=0.33\textwidth,height=4cm]{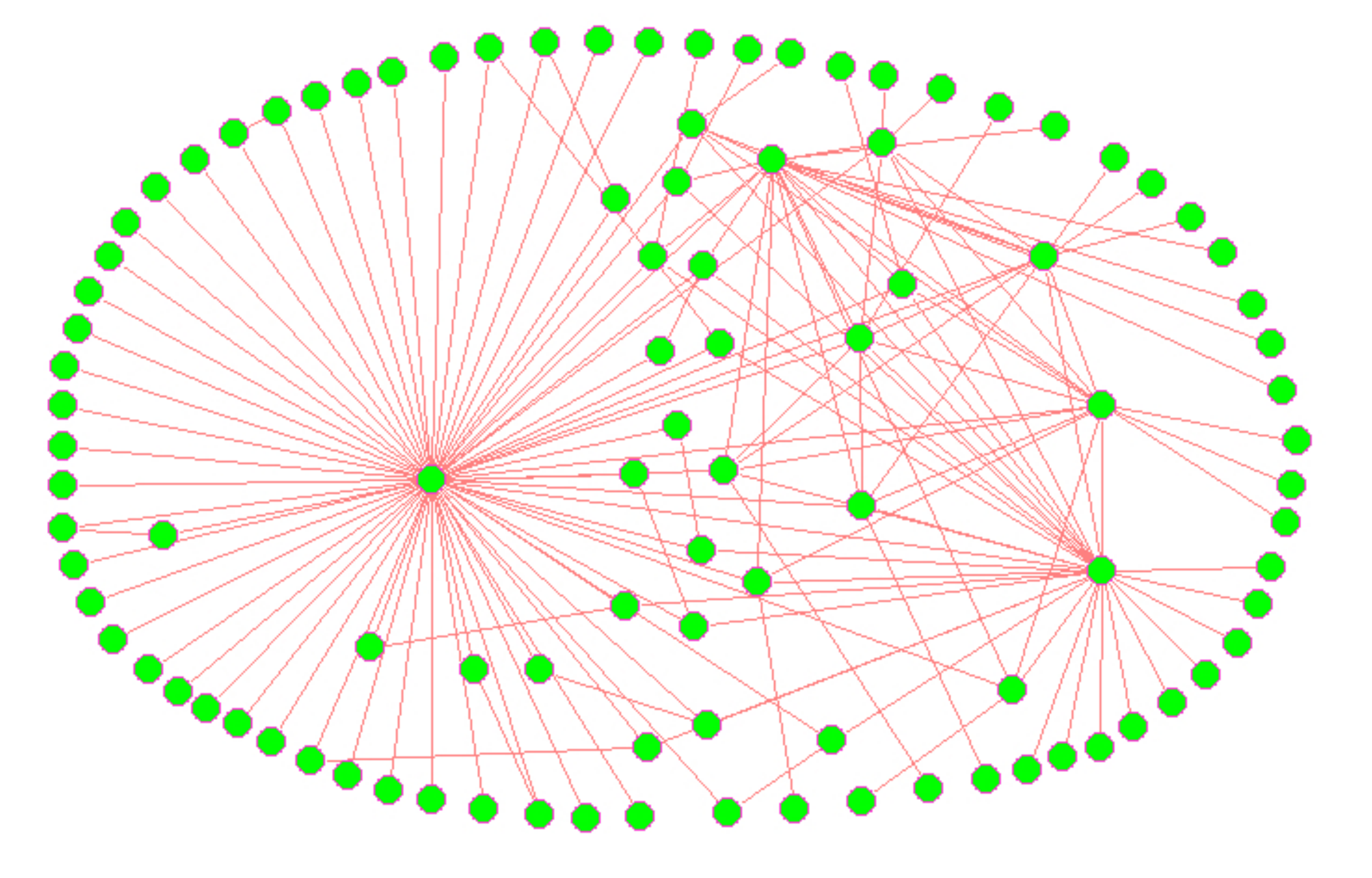}\label{a}}
	\subfigure[$SF_3$]{\includegraphics[width=0.33\textwidth,height=4cm]{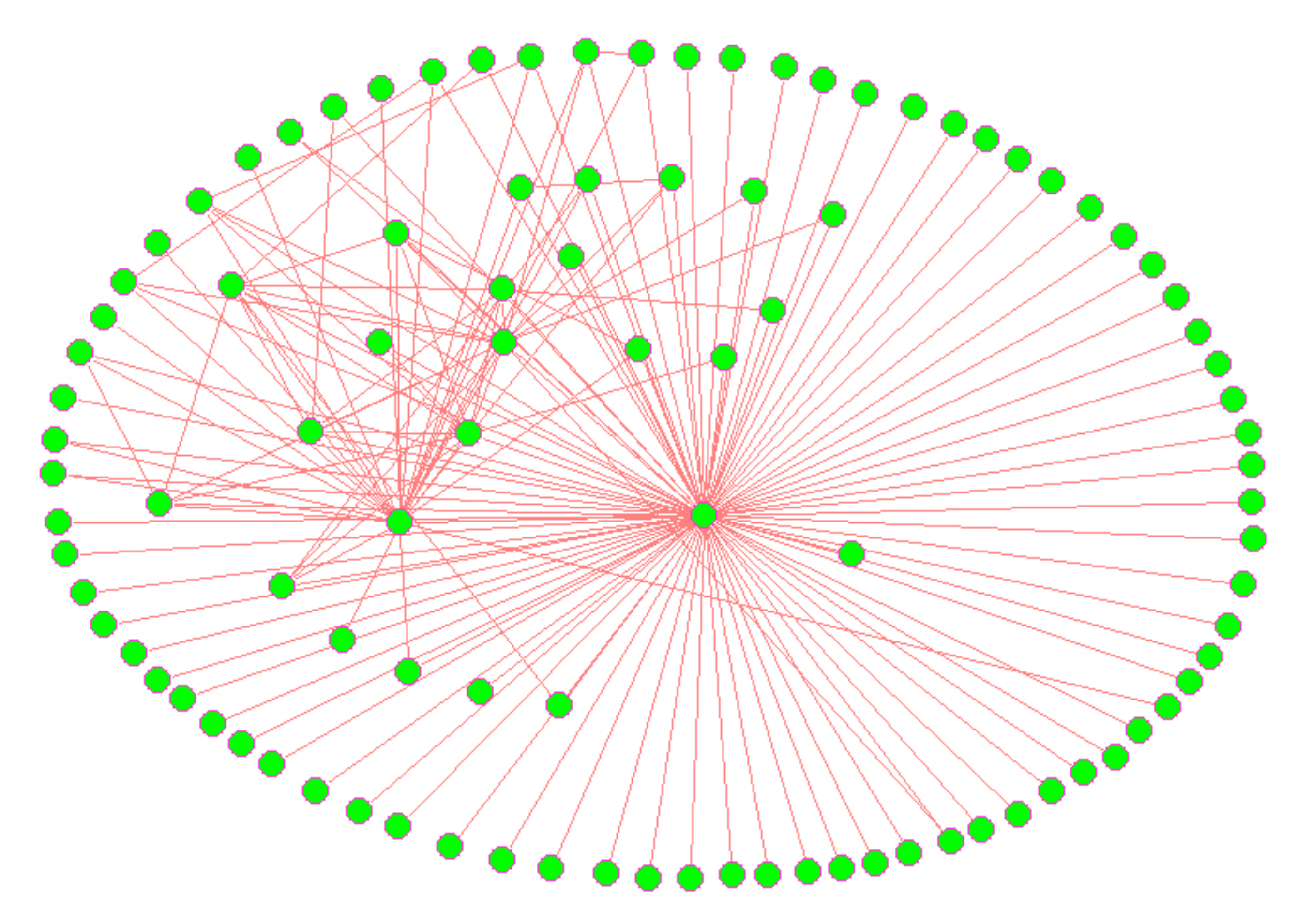}\label{a}}
	\vspace{-3ex}
	\caption{Three scale-free networks with 100 nodes, 162 edges, and power exponents $r$ = 1.9, 2.0, and 2.1, respectively.}
\end{figure}

\section{The dynamics of the DPA model}

The DPA model has significant influence on the performance of the IBD strategy. This section aims to experimentally reveal the influence of different factors on the dynamics of the DPA model.

\subsection{The influence of the four basic parameters}

%For the purpose of examining the influence of the three basic parameters in the DPA model on the dynamics of the DPA model, take the small-world network in Fig. 2(a) as the WOM network, and choose the top $5\%$ highest-degree nodes as the seeds.

This subsection examines the influence of the four basic factors (the WOM force, the rigid demand, the lure force and the viscosity) on the dynamics of the DPA model, respectively.

Comprehensive experiments show that, typically, the influence of the WOM force on the dynamics of the DPA is as shown in Fig. 4. In general, the following conclusions are drawn.

\begin{enumerate}
	\item[(a)] For any WOM force, the expected fractions of potential and adopting customers level off, respectively.
	\item[(b)] With the rise of the WOM force, the steady expected fractions of potential and adopting customers go up, respectively.
\end{enumerate}

\begin{figure}[H]
   \subfigure{\includegraphics[width=0.5\textwidth,height=5cm]{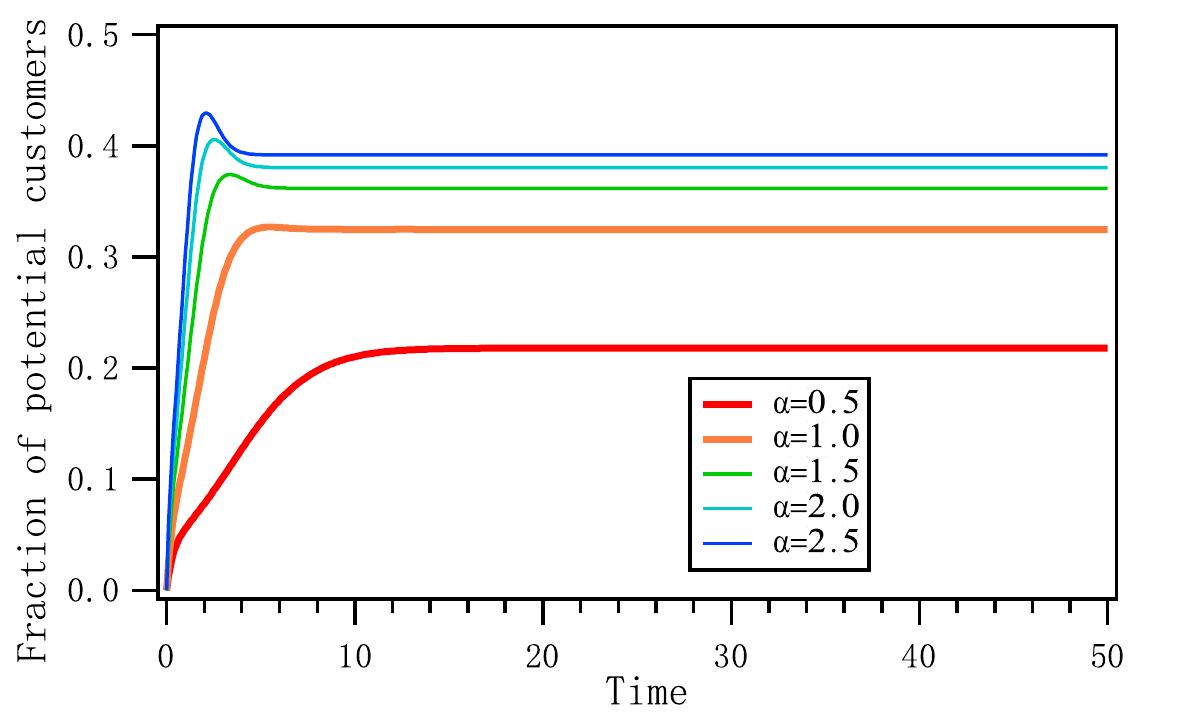}
   \label{fig:a} }
   \subfigure{\includegraphics[width=0.5\textwidth,height=5cm]{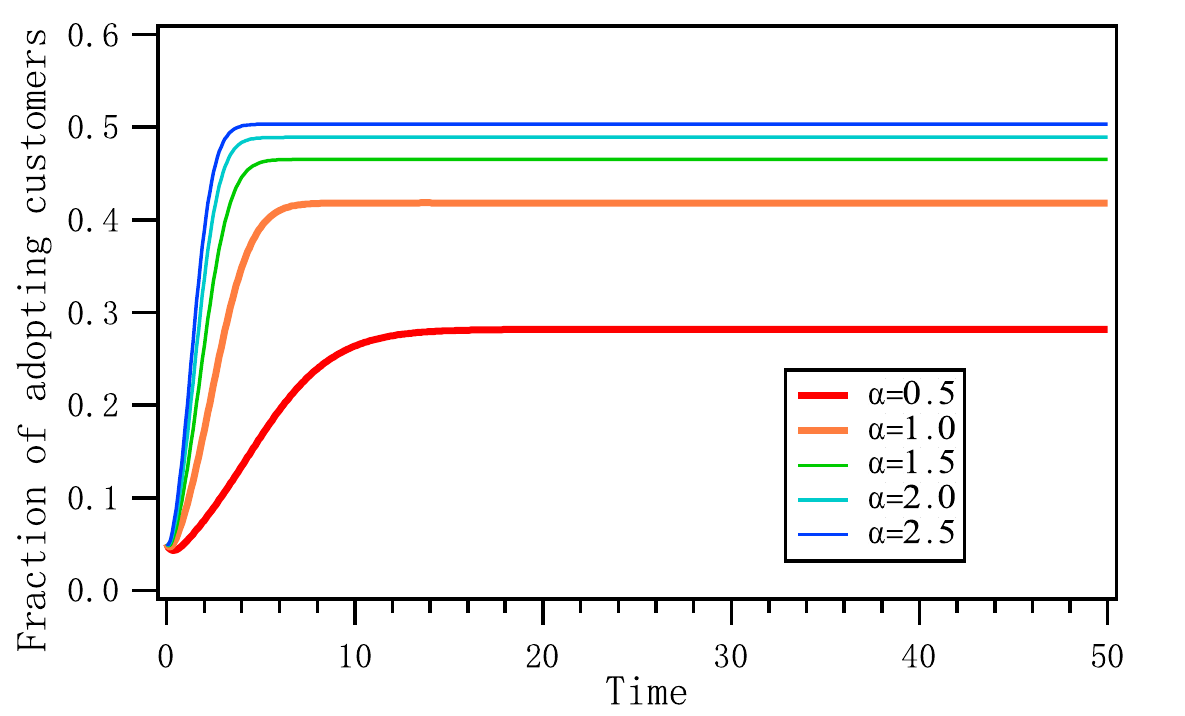}
   \label{fig:b} }
   \vspace{-5ex}
   \caption{The influence of the WOM force on the dynamics of the DPA model.}
\end{figure}

Comprehensive experiments show that, typically, the influence of the rigid demand on the dynamics of the DPA model is as shown in Fig. 5. In general, the following conclusions are drawn.

\begin{enumerate}
	\item[(a)] For any rigid demand, the expected fractions of potential and adopting customers level off, respectively.
	\item[(b)] With the rise of the rigid demand, the steady expected fraction of potential customers goes down.
	\item[(c)] With the rise of the rigid demand, the steady expected fraction of adopting customers goes up.
\end{enumerate}

\begin{figure}[H]
	\subfigure{\includegraphics[width=0.5\textwidth,height=5cm]{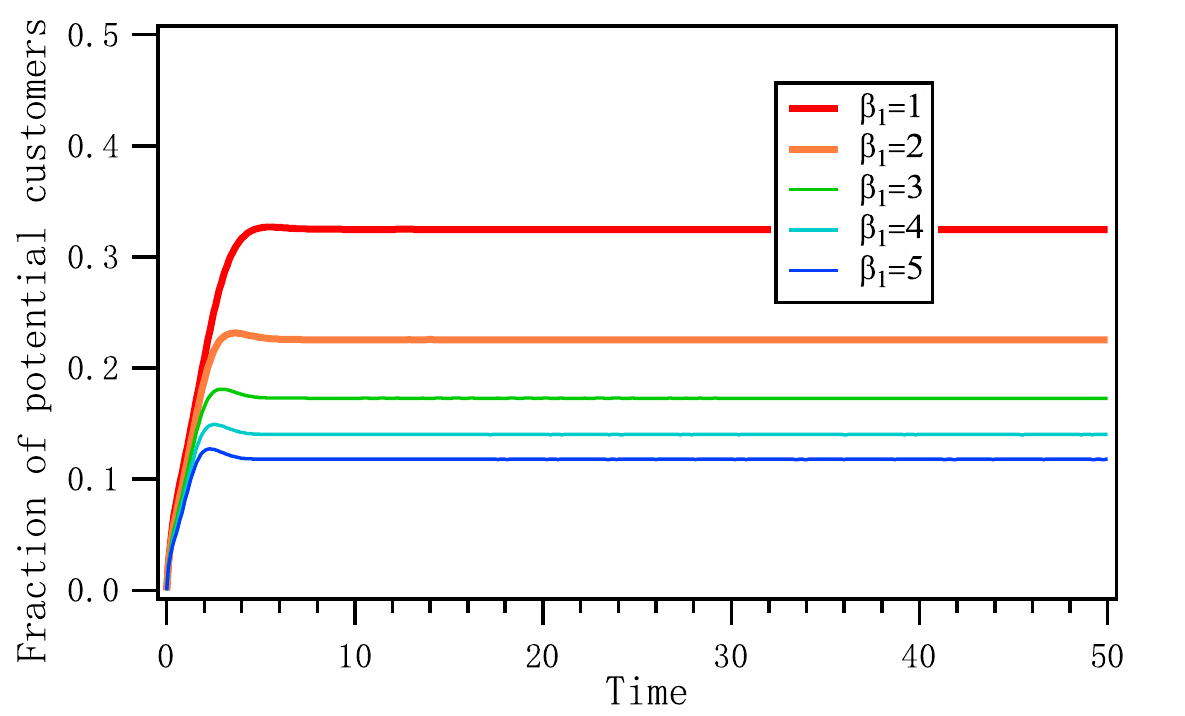}
		\label{fig:a} }
	\subfigure{\includegraphics[width=0.5\textwidth,height=5cm]{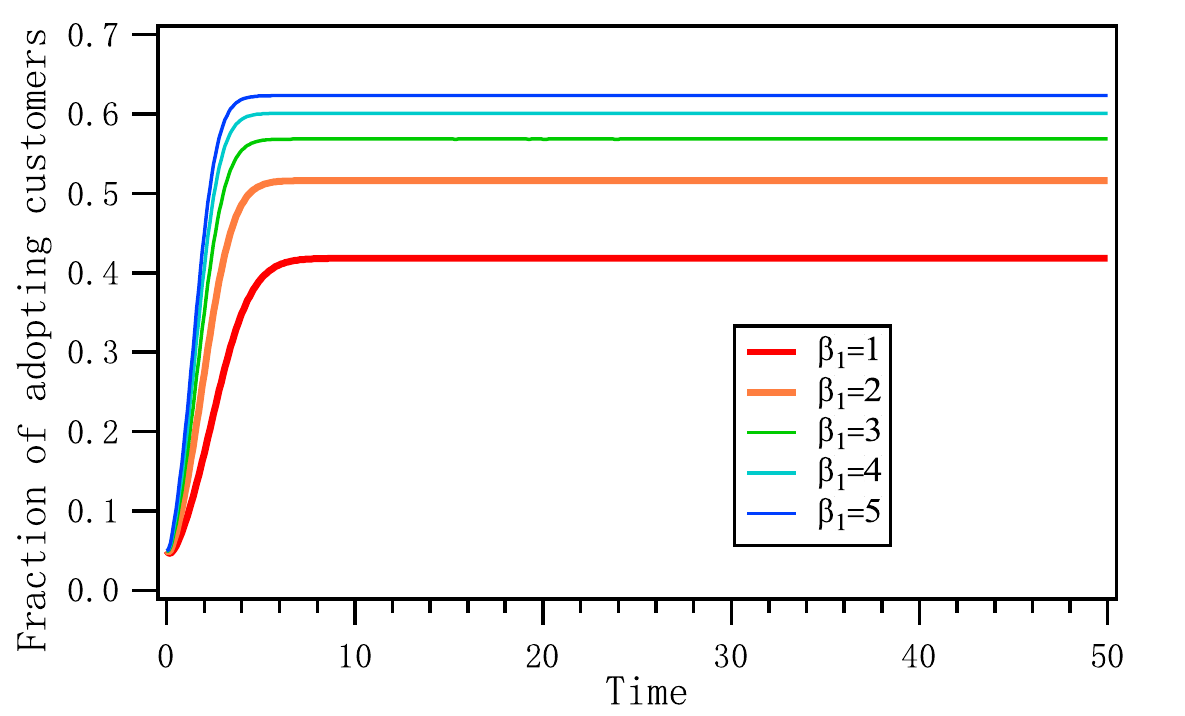}
		\label{fig:b} }
	\vspace{-5ex}
	\caption{The influence of the rigid demand on the dynamics of the DPA model.}
\end{figure}

Comprehensive experiments show that, typically, the influence of the lure force on the dynamics of the DPA model is as shown in Fig. 6. In general, the following conclusions are drawn.

\begin{enumerate}
	\item[(a)] For any lure force, the expected fractions of potential and adopting customers level off, respectively.
	\item[(b)] With the rise of the lure force, the steady expected fraction of potential customers goes down.
	\item[(c)] With the rise of the lure force, the steady expected fraction of adopting customers goes up.
\end{enumerate}

\begin{figure}[H]
   \subfigure{\includegraphics[width=0.5\textwidth,height=5cm]{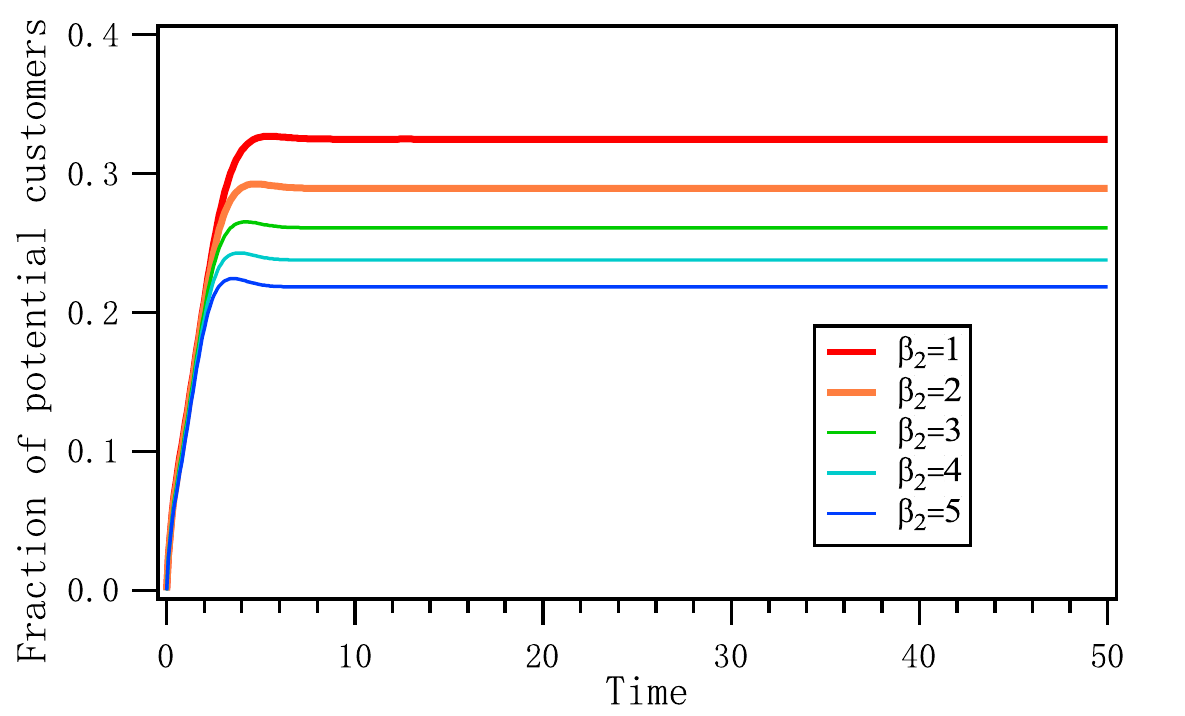}
   \label{fig:a} }
   \subfigure{\includegraphics[width=0.5\textwidth,height=5cm]{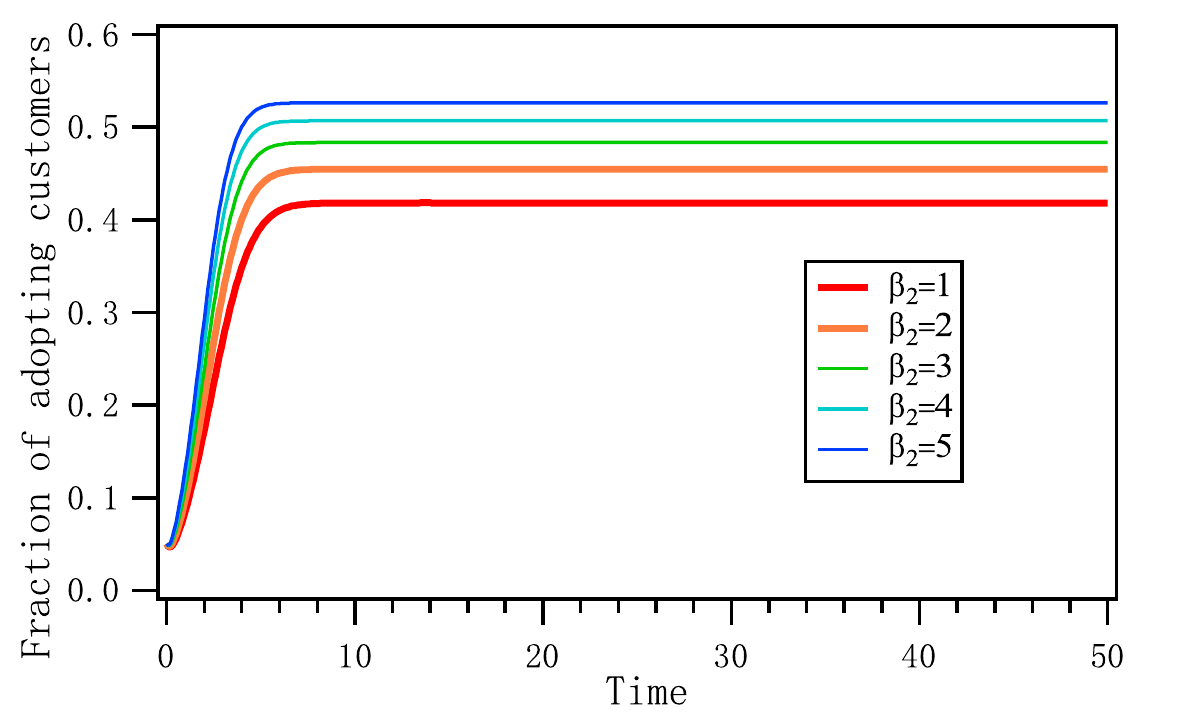}
   \label{fig:b} }
   \vspace{-5ex}
   \caption{The influence of the lure force on the dynamics of the DPA model.}
\end{figure}

Comprehensive experiments show that, typically, the influence of the viscosity on the dynamics of the DPA model is as shown in Fig. 7. In general, the following conclusions are drawn.

\begin{enumerate}
	\item[(a)] For any viscosity, the expected fractions of potential and adopting customers level off, respectively.
	\item[(b)] With the rise of the viscosity, the steady expected fraction of adopting customers go down.
	\item[(c)] It is unclear how the steady expected fraction of potential customers is dependent on the viscosity.
\end{enumerate}

\begin{figure}[H]
   \subfigure{\includegraphics[width=0.5\textwidth,height=5cm]{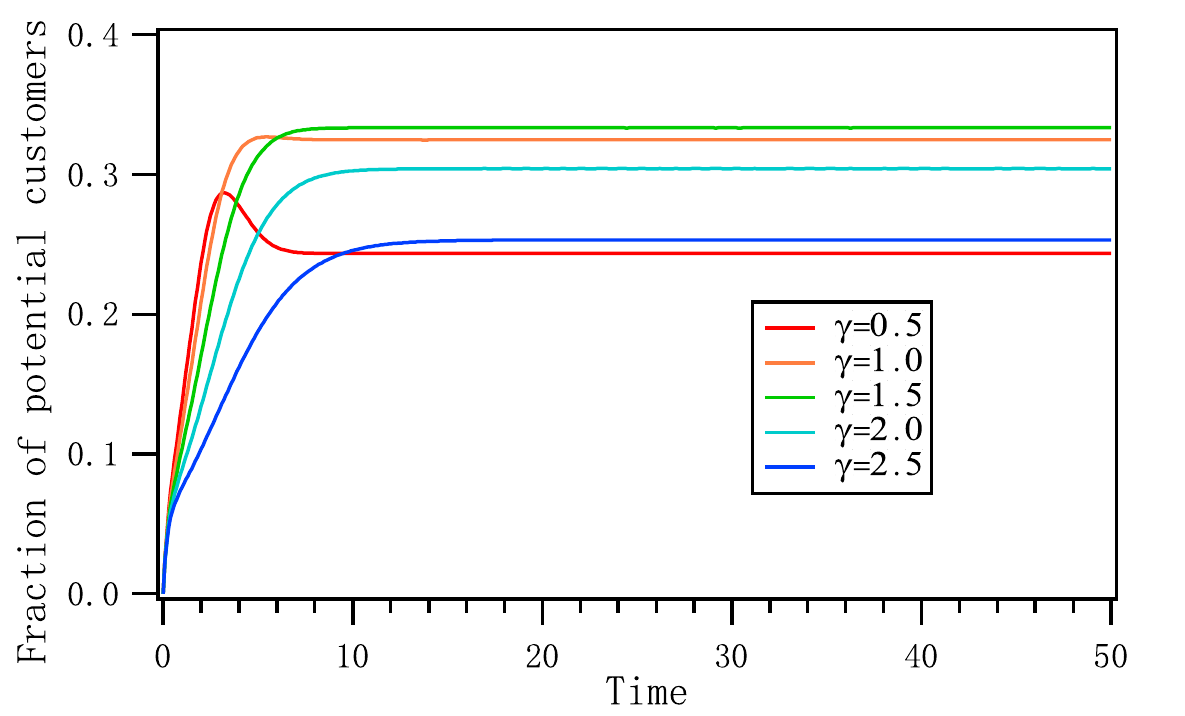}
   \label{fig:a} }
   \subfigure{\includegraphics[width=0.5\textwidth,height=5cm]{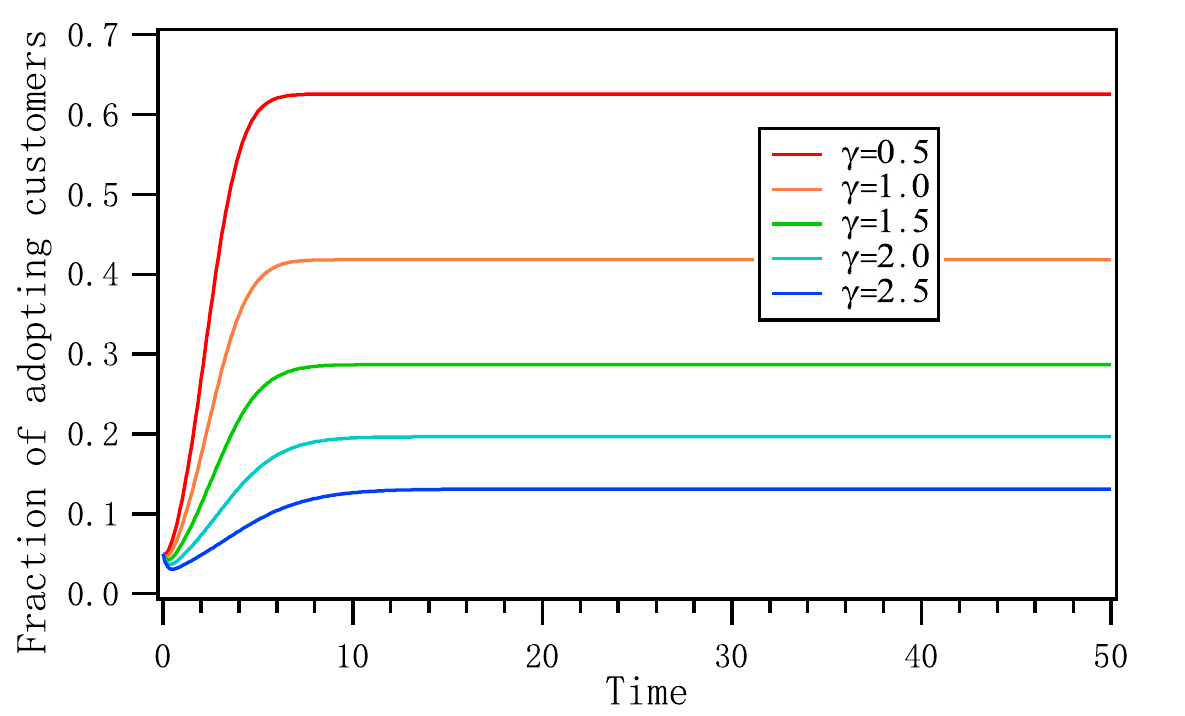}
   \label{fig:b} }
   \vspace{-5ex}
   \caption{The influence of the viscosity on the dynamics of the DPA model.}
\end{figure}

\subsection{The influence of the structure of the WOM network}

Comprehensive experiments show that, typically, the influence of the randomness of the small-world WOM network on the dynamics of the DPA model is as shown in Fig. 8. In generally, the following conclusions are drawn.

\begin{enumerate}
	\item[(a)] For any small-world WOM network, the expected fractions of potential and adopting customers level off, respectively.
	\item[(b)] With the rise of the randomness of the small-world WOM network, the steady expected fraction of potential customers goes up.
	\item[(c)] With the rise of the randomness of the small-world WOM network, the steady expected fraction of adopting customers goes down.
\end{enumerate}

\begin{figure}[H]
   \subfigure{\includegraphics[width=0.5\textwidth,height=5cm]{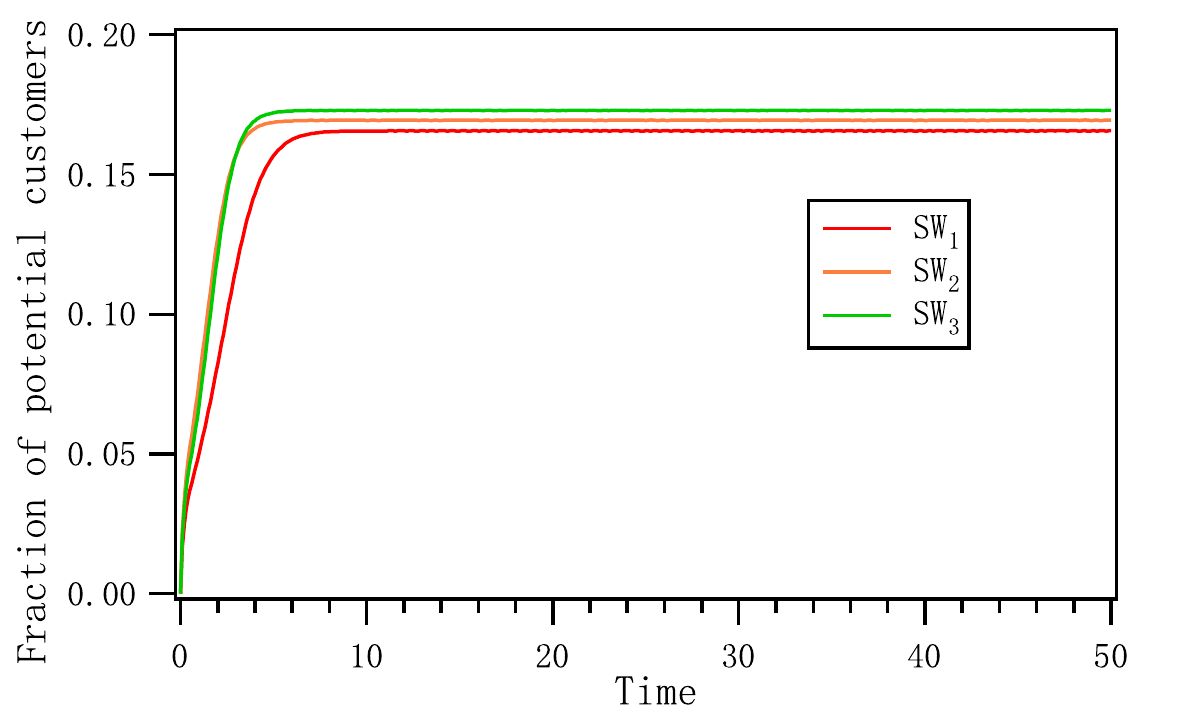}
   \label{fig:a} }
   \subfigure{\includegraphics[width=0.5\textwidth,height=5cm]{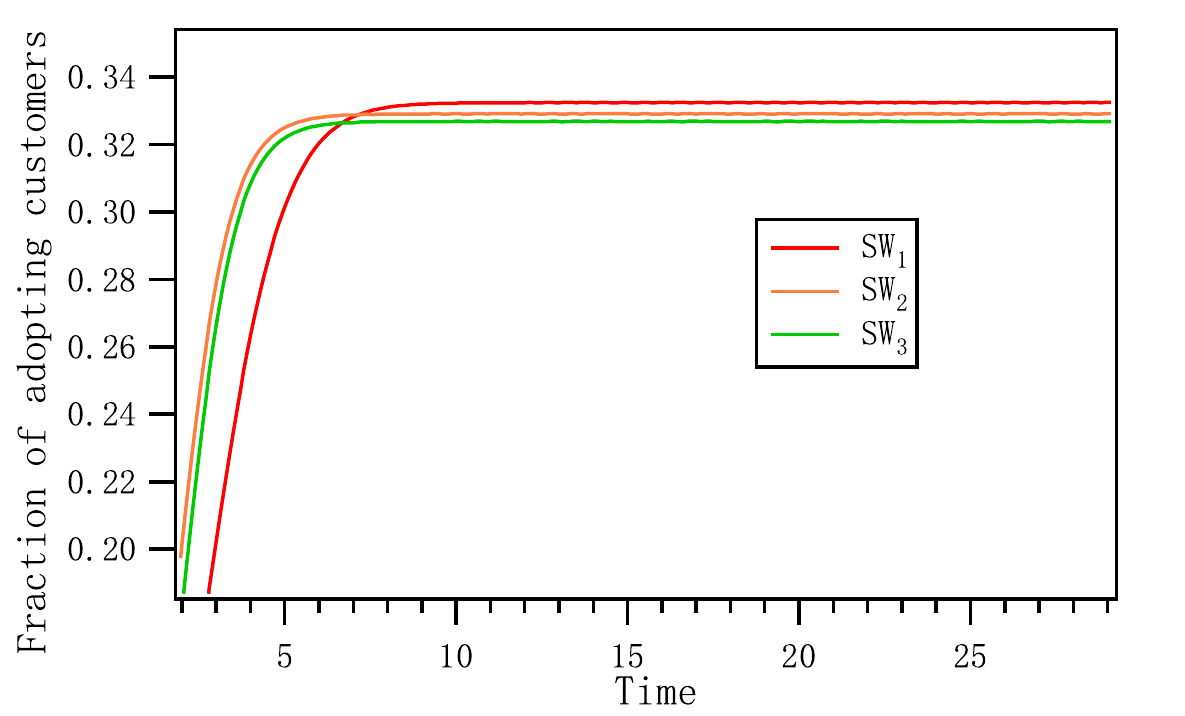}
   \label{fig:b} }
   \vspace{-5ex}
   \caption{The influence of the randomness of the small-world WOM network on the dynamics of the DPA model.}
\end{figure}

Comprehensive experiments show that, typically, the influence of the heterogeneity of the scale-free WOM network on the dynamics of the DAP model is as shown in Fig. 9. In generally, the following conclusions are drawn.

\begin{enumerate}
	\item[(a)] For any scale-free WOM network, the expected fractions of potential and adopting customers level off, respectively.
	\item[(b)] With the rise of the heterogeneity of the scale-free WOM network, the steady expected fraction of potential and adopting customers goes up.
\end{enumerate}

\begin{figure}[H]
   \subfigure{\includegraphics[width=0.5\textwidth,height=5cm]{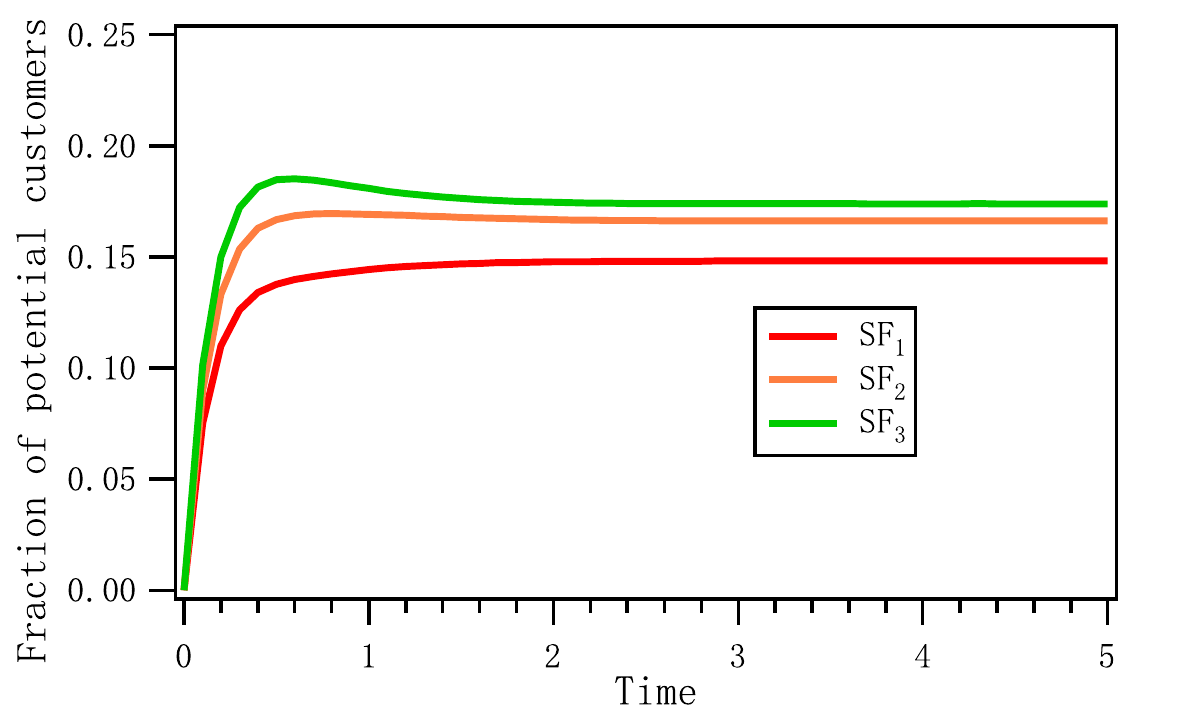}
   \label{fig:a} }
   \subfigure{\includegraphics[width=0.5\textwidth,height=5cm]{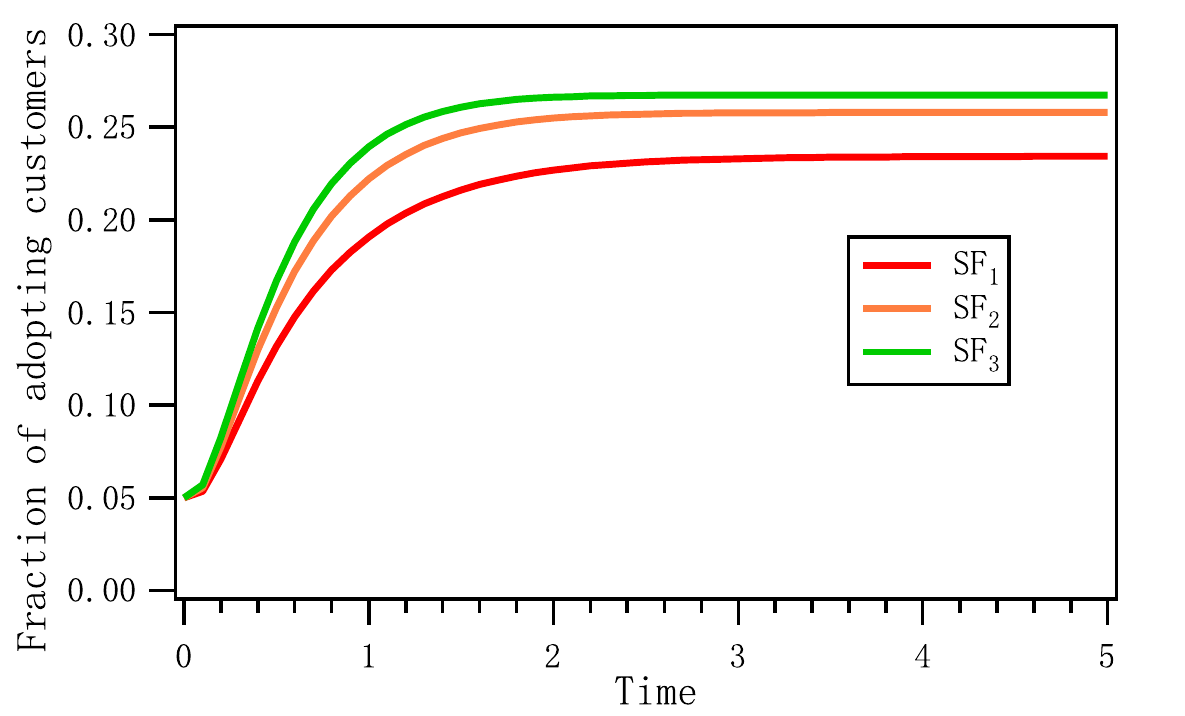}
   \label{fig:b} }
   \vspace{-5ex}
   \caption{The influence of the structure of a scale-free network on the dynamics of the DPA model.}
\end{figure}

\subsection{The influence of the basic discount}

Comprehensive experiments show that, typically, the influence of the basic discount on the dynamics of the DPA model is as shown in Fig. 10. In generally, the following conclusions are drawn.

\begin{enumerate}
	\item[(a)] For any basic discount, the expected fractions of potential and adopting customers level off, respectively.
	\item[(b)] With the rise of the basic discount, the steady expected fraction of potential customers goes down.
	\item[(c)] With the rise of the basic discount, the steady expected fraction of adopting customers goes up.
\end{enumerate}

\begin{figure}[H]
   \subfigure{\includegraphics[width=0.5\textwidth,height=5cm]{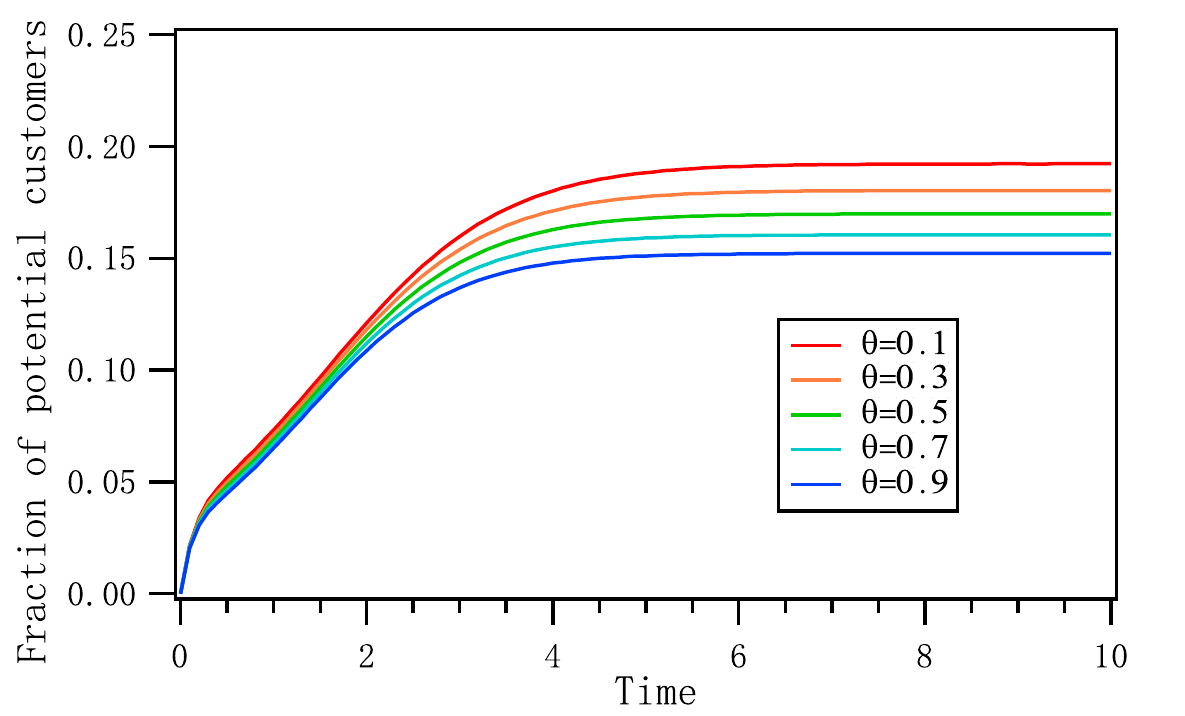}
   \label{fig:a} }
   \subfigure{\includegraphics[width=0.5\textwidth,height=5cm]{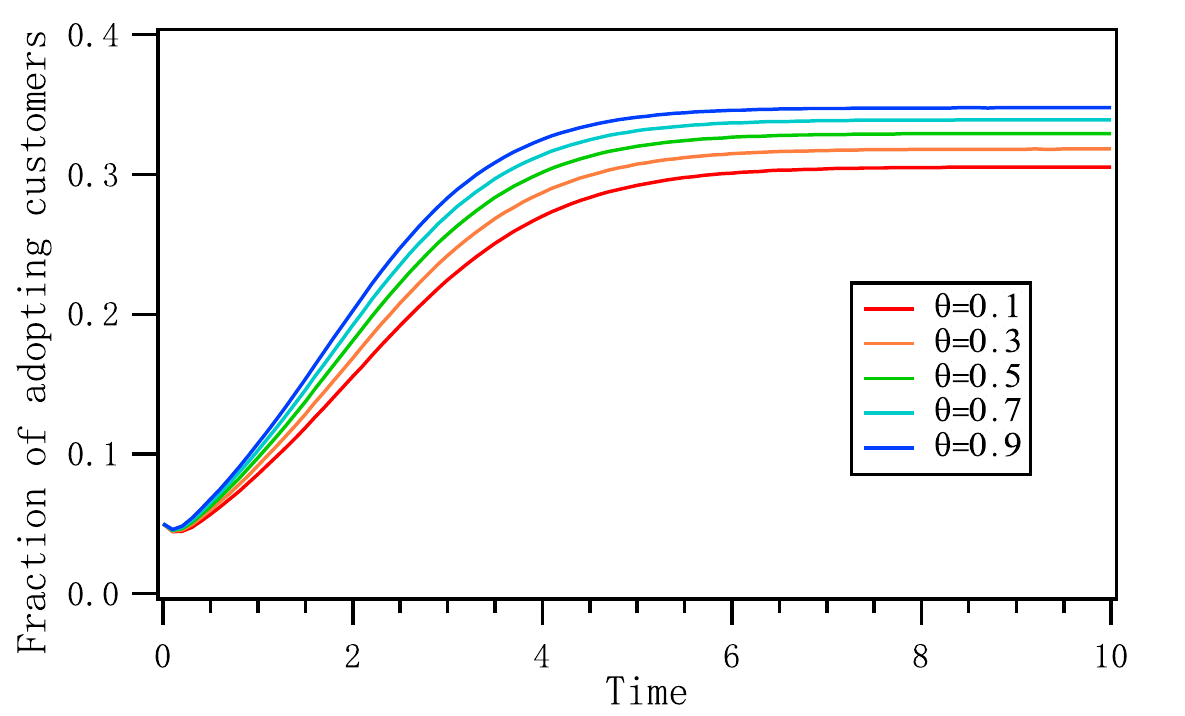}
   \label{fig:b} }
   \vspace{-5ex}
   \caption{The influence of the basic discount on the dynamics of the DPA model.}
\end{figure}

\section{The expected profit of the IBD strategy}

This section aims to reveal the influence of different factors on the expected profit of the IBD strategy.

\subsection{The influence of the four basic parameters}

Comprehensive experiments show that, typically, the influence of the WOM force on the expected profit of the IBD strategy is as shown in Fig. 11. In general, it is concluded that, with the rise of the WOM force, the expected profit undergoes an S-shaped growth: it grows very slowly at the first stage, takes off at the intermediate stage, and flatten out at the final stage. In marketing practice, the promoted products must be of high quality and with good experience so as to enhance the WOM force.
\begin{figure}[H]
  \centering
  \includegraphics[width=8cm]{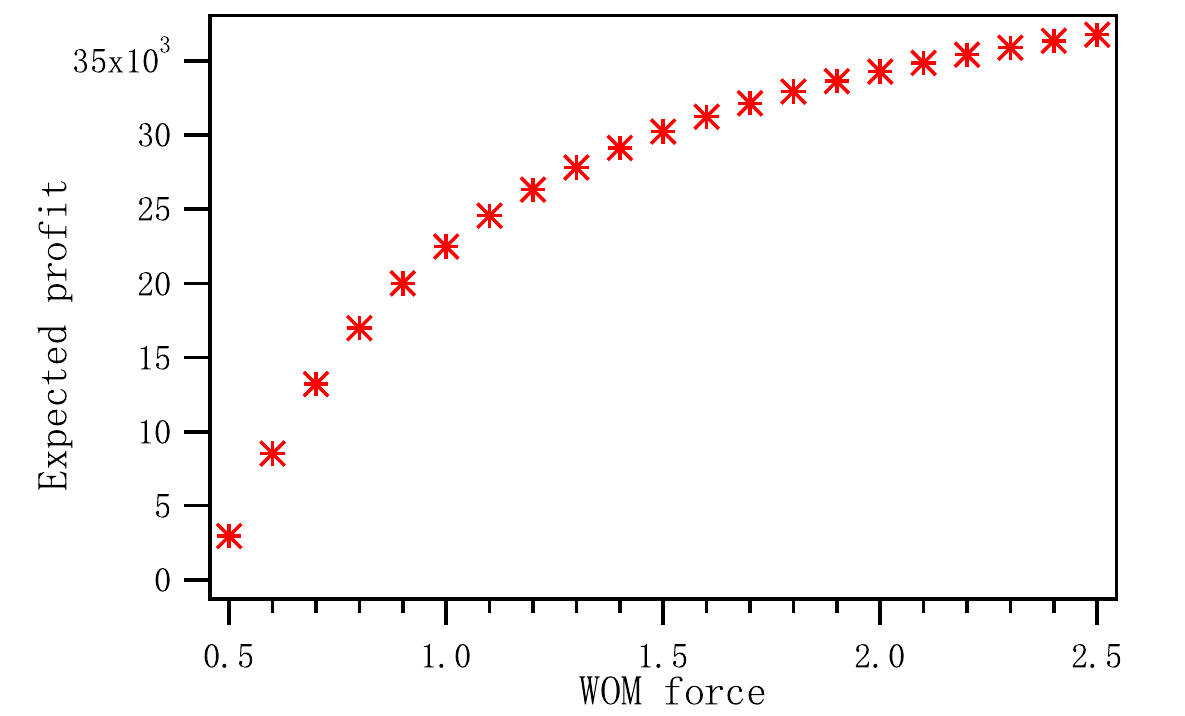}\\
  \caption{The expected profit versus the WOM force.}
\end{figure}

Comprehensive experiments show that, typically, the influence of the rigid demand on the expected profit of the IBD strategy is as shown in Fig. 12. In general, it is concluded that, with the rise of the rigid demand, the expected profit grows, but the growth rate slows down.

\begin{figure}[H]
  \centering
  \includegraphics[width=8cm]{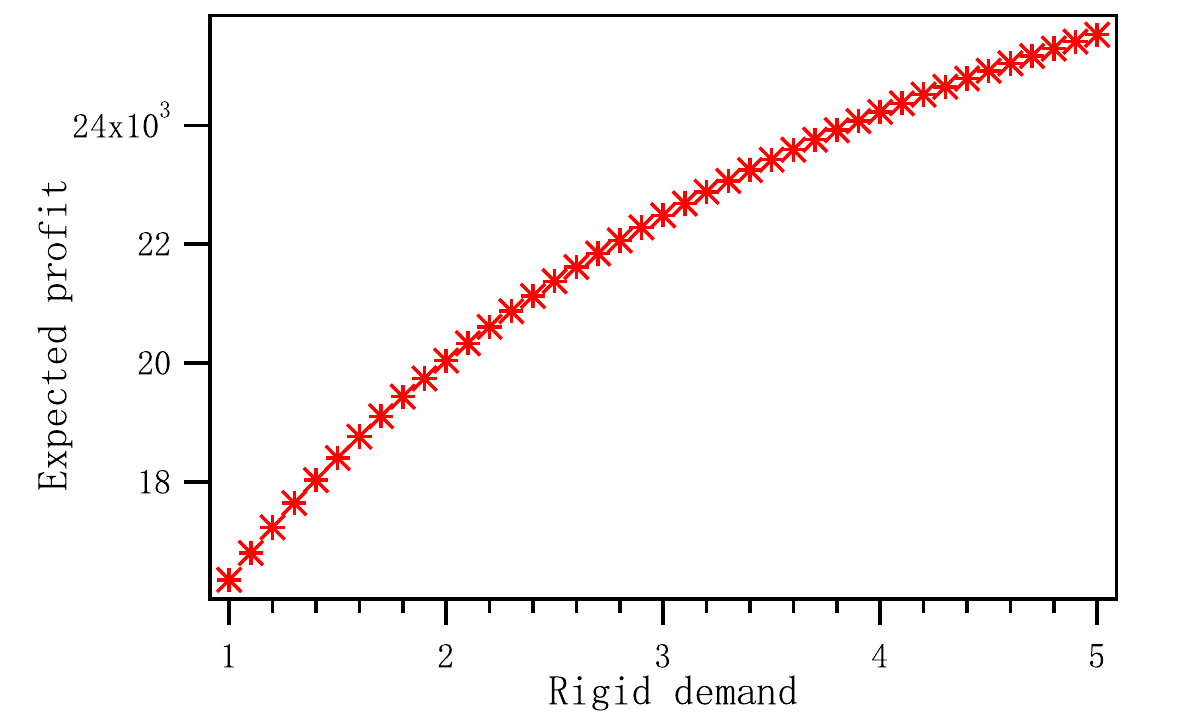}\\
  \caption{The expected profit versus the rigid demand.}
\end{figure}

Comprehensive experiments show that, typically, the influence of the lure force on the expected profit of the IBD strategy is as shown in Fig. 13. In general, it is concluded that, with the rise of the lure force, the expected profit grows, but the growth rate slows down. In marketing practice, promotional measures such as providing small gift or offering the next-campaign coupon should be taken so as to intensify the lure force.

\begin{figure}[H]
  \centering
  \includegraphics[width=8cm]{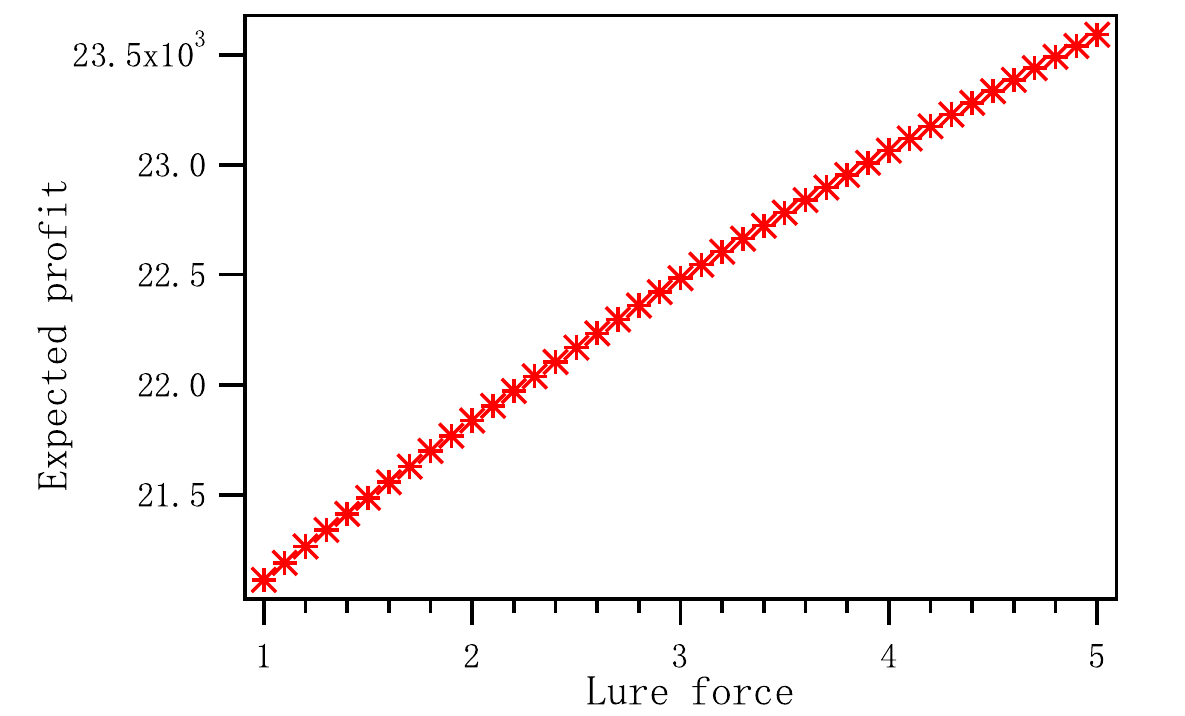}\\
  \caption{The expected profit versus the lure force.}
\end{figure}

Comprehensive experiments show that, typically, the influence of the viscosity on the expected profit is as shown in Fig. 14. In general, it is concluded that, with the rise of the viscosity force, the expected profit goes up first, then goes down. In marketing practice, measures such as pushing new discount information should be taken so as to enhance the viscosity to the turning point.

\begin{figure}[H]
  \centering
  \includegraphics[width=8cm]{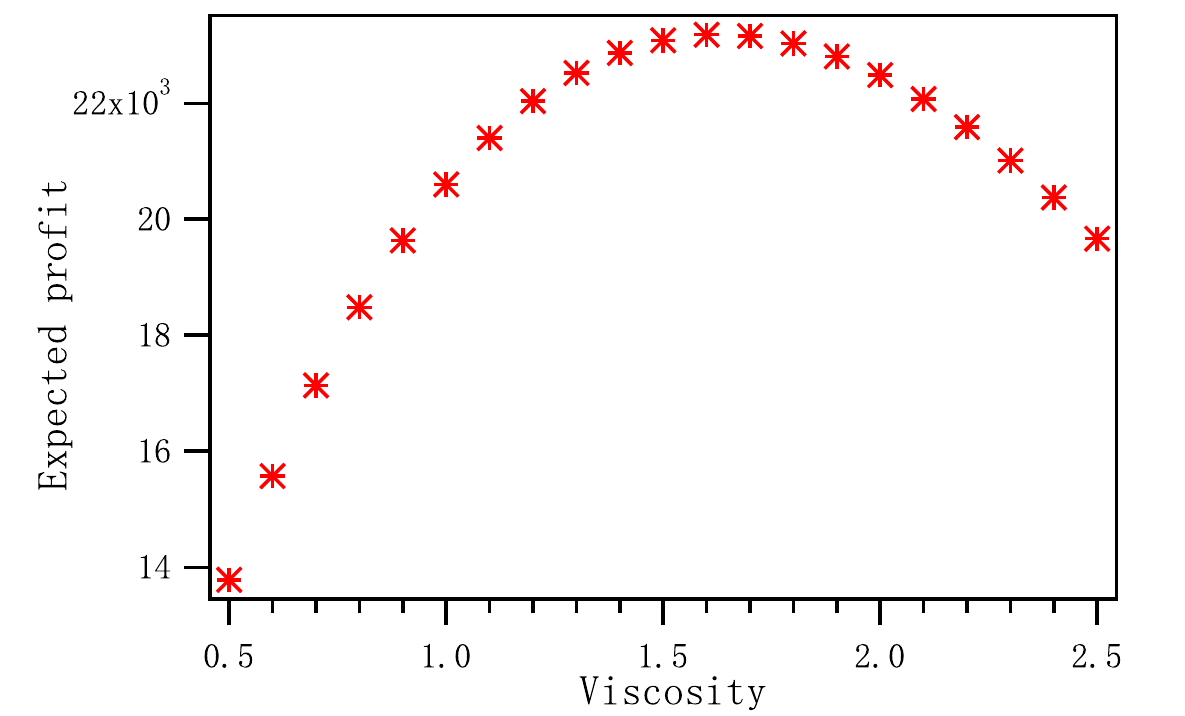}\\
  \caption{The expected profit versus the viscosity.}
\end{figure}

\subsection{The influence of the WOM network}

Comprehensive experiments show that, typically, the influence of the randomness of the small-world WOM network on the expected profit of the IBD strategy is as shown in Fig. 15. It is concluded that there is a threshold of $\beta_1$ such that (1) when $\beta_1$ is lower than the threshold, the expected profit goes down with the rise of the randomness of the small-world network, and (2) when $\beta_1$ exceeds the threshold, the expected profit goes up with the rise of the randomness of the small-world network.

\begin{figure}[H]
   \subfigure[$\beta_1=0.1$]{\includegraphics[width=0.33\textwidth]{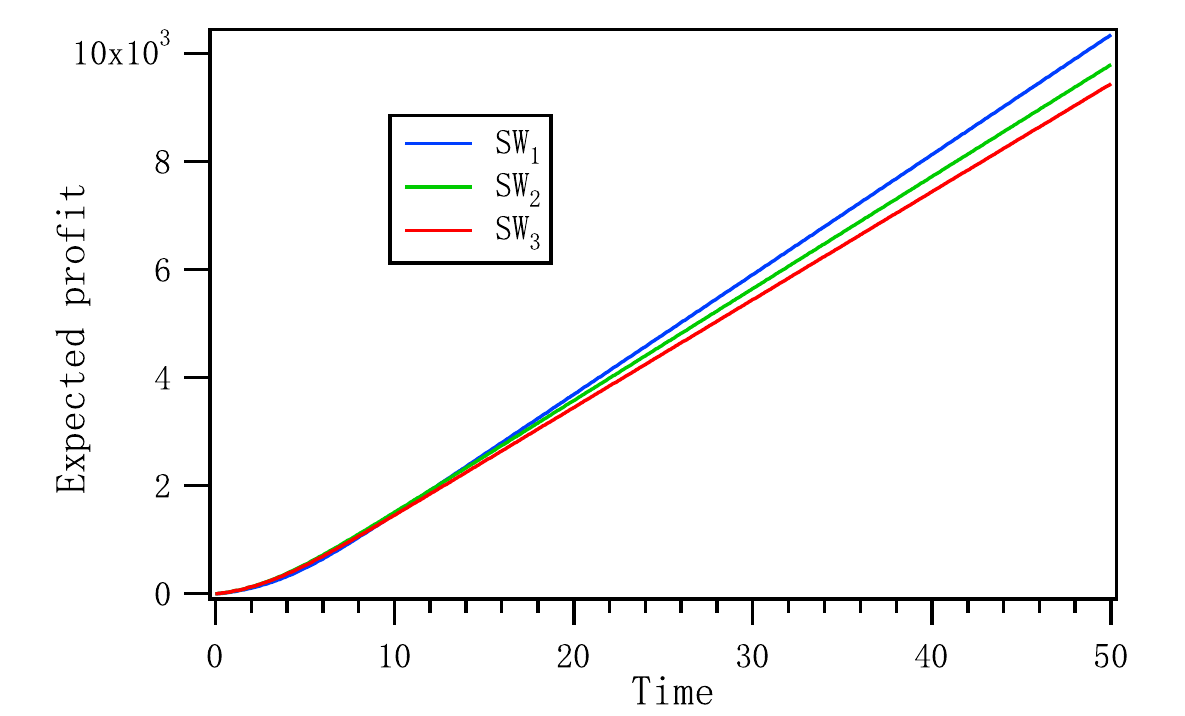}
   \label{fig:a} }
   \subfigure[$\beta_1=0.3$]{\includegraphics[width=0.33\textwidth]{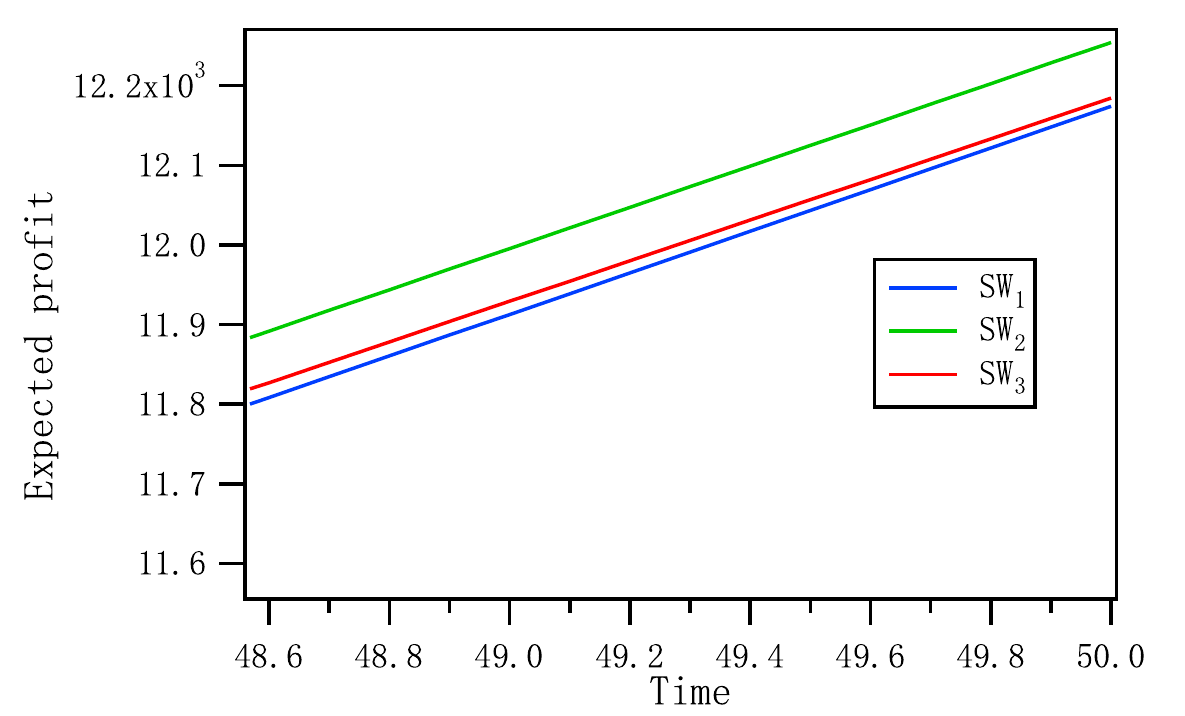}
   \label{fig:b} }
   \subfigure[$\beta_1=1$]{\includegraphics[width=0.33\textwidth]{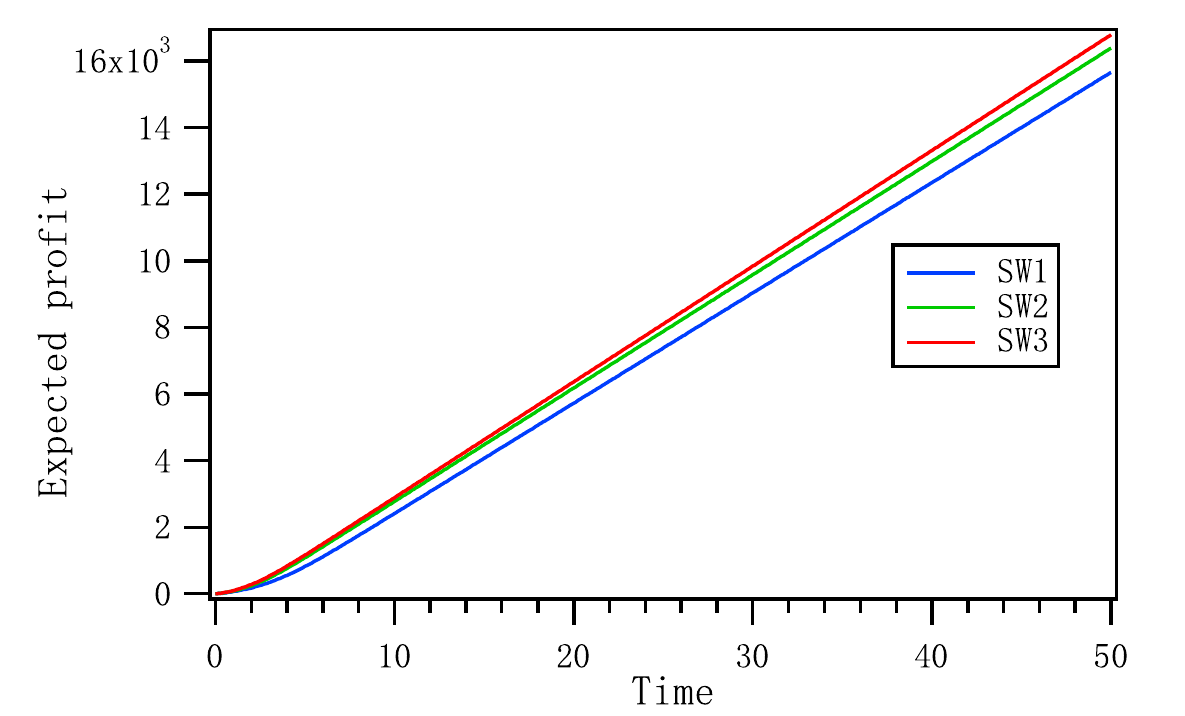}
   \label{fig:b} }
   \vspace{-2ex}
   \caption{The expected profit versus the randomness of the small-world WOM network.}
\end{figure}

Comprehensive experiments show that, typically, the influence of the heterogeneity of the scale-free WOM network on the expected profit of the IBD strategy is as shown in Fig. 16. It is concluded that there is a threshold of $\beta_1$ such that (1) when $\beta_1$ is lower than the threshold, the expected profit goes down with the rise of the heterogeneity of the scale-free network, and (2) when $\beta_1$ exceeds the threshold, the expected profit goes up with the rise of the heterogeneity of the scale-free network.

\begin{figure}[H]
   \subfigure[$\beta_1=0.05$]{\includegraphics[width=0.33\textwidth]{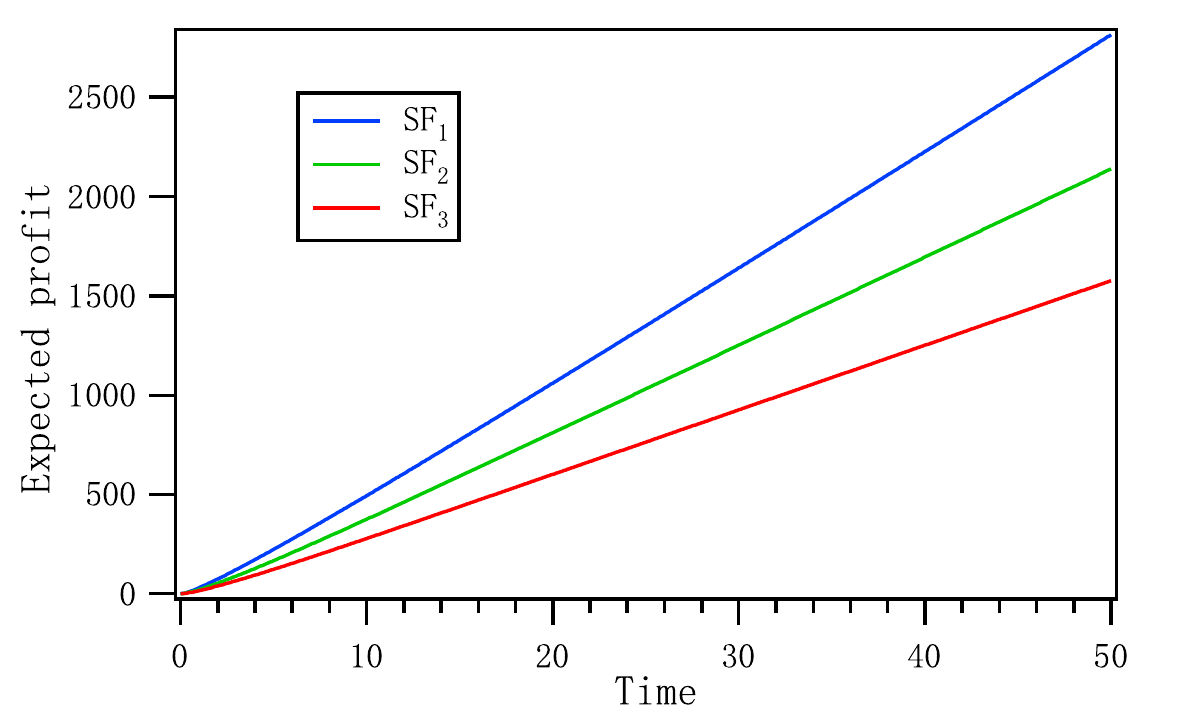}
   \label{fig:a} }
   \subfigure[$\beta_1=0.12$]{\includegraphics[width=0.33\textwidth]{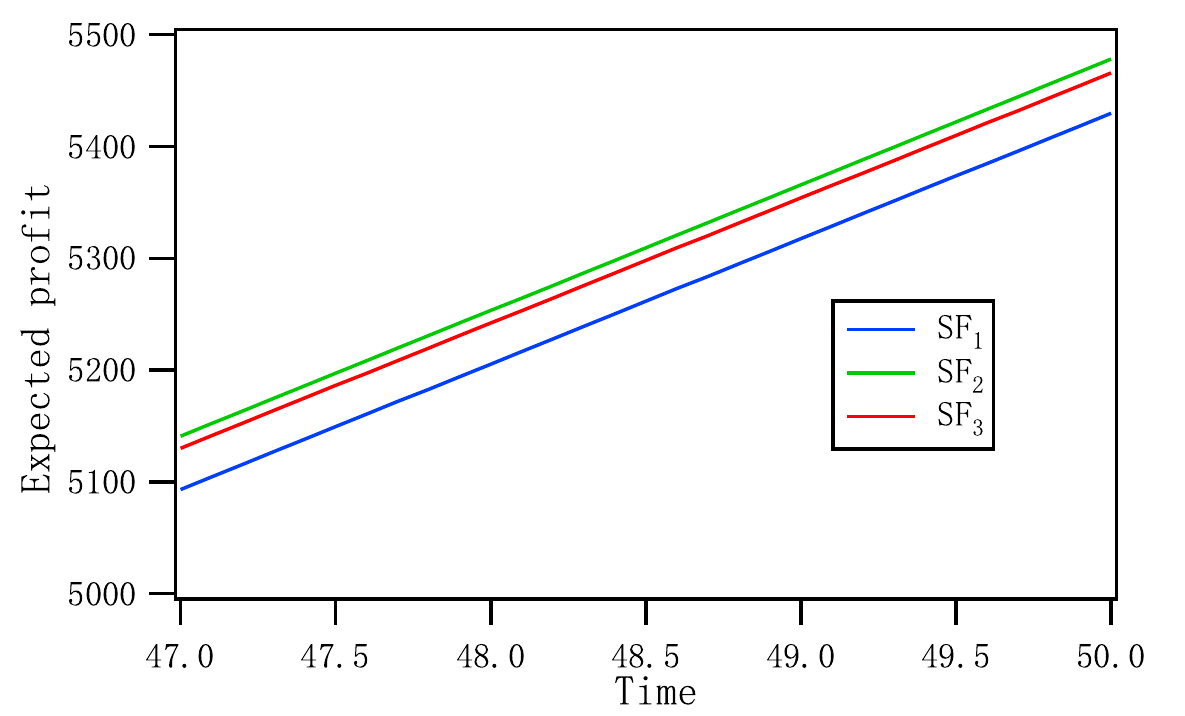}
   \label{fig:b} }
   \subfigure[$\beta_1=1$]{\includegraphics[width=0.33\textwidth]{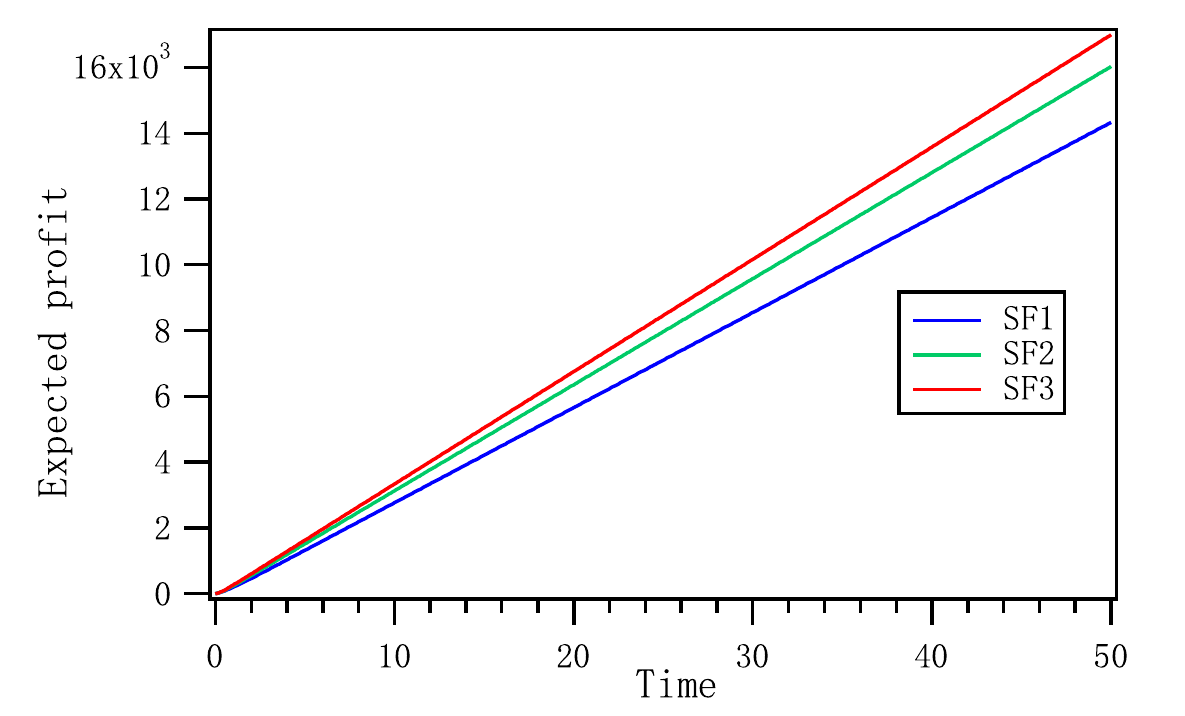}
   \label{fig:b} }
   \vspace{-2ex}
   \caption{The expected profit versus the heterogeneity of the scale-free WOM network.}
\end{figure}

\subsection{The influence of the basic discount}

Comprehensive experiments show that, typically, the influence of the basic discount on the expected profit of the IBD strategy is as shown in Fig. 17. In general, it is concluded that there is a threshold of the basic discount such that (1) when the basic disount is lower than the threshold, the expected profit grows with the rise of the basic discount, and (2) when the basic disount exceeds the threshold, the expected profit drops with the rise of the basic discount. This phenomenon can be explained as follows. When the basic discount is very lower, the marketing campaign is less attractive, leading to a low marketing profit. Whereas when the basic discount is very high, the high gross profit is severely diminished by the high discounts, also leading to a low marketing profit. In marketing practice, an in-between basic discount would help achieve the maximum possible profit.

\begin{figure}[H]
  \centering
  \includegraphics[width=8cm]{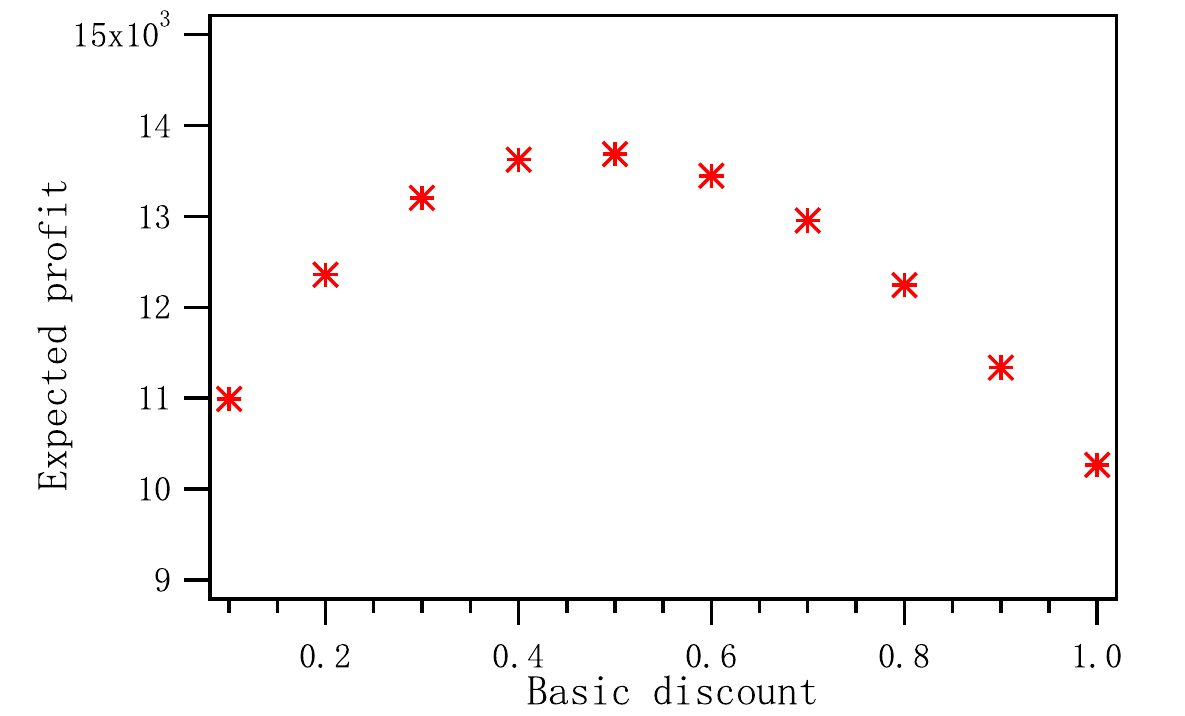}\\
  \caption{The expected profit versus basic discount.}
\end{figure}

\section{Conclusions and remarks}

In the context of WOM marketing, a natural discount strategy has been proposed. By modeling the marketing profit, the performance of the discount strategy has been evaluated. On this basis, some promotional measures have been recommended.

Towards this direction, there are many problems to be solved. The influential degree in the proposed discount strategy may be replaced with other influence indexes \cite{Sabidussi1966, Freeman1979, Bonacich2007, LuLY2011, ChenDB2012, ChenDB2013}, resulting in many similar discount strategies. A comparative study can be conducted so as to find out the best one. For the purpose of comprehensively evaluating the performance of the proposed discount strategy, a profit model that allows consumers to purchase more than one item each time must be developed. In order to further enhance the marketing profit, the static discount strategy must be made dynamic and more flexible \cite{Preciado2014, Shakeri2015, YangLX2016, Nowzari2016, ZhangTR2017}.

\section*{Acknowledgments}

%The authors are grateful to the three anonymous reviewers for their valuable comments and suggestions.

This work is supported by Natural Science Foundation of China (Grant Nos. 61572006, 71301177), Sci-Tech Support Program of China (Grant
No. 2015BAF05B03), Basic and Advanced Research Program of
Chongqing (Grant No. cstc2013jcyjA1658) and Fundamental Research Funds for the Central Universities (Grant No. 106112014CDJZR008823).

%% References

\section*{References}
\bibliographystyle{elsarticle-num}
\bibliography{<your-bib-database>}



\end{document}